\documentclass{aimscleaned} 

\usepackage{amsmath}
\usepackage{xfrac}
\usepackage{graphics}
\usepackage{graphicx}
\usepackage[dvipsnames]{xcolor}
\usepackage[xcolor]{mdframed}
\usepackage{epstopdf}
\usepackage{dsfont}
\usepackage{enumitem}

\usepackage{caption}
\usepackage{tikz}
\usetikzlibrary{decorations.pathreplacing,automata,positioning,arrows}

\newmdenv[
	backgroundcolor=yellow!20,
	leftline=false,
	rightline=false,
	bottomline=false,
	linewidth=3pt,
	linecolor=ForestGreen
]{myframe}

\newcommand{\one}{\mathds{1}}
\newcommand{\RSSI}{\mathcal{R}}

\graphicspath{{./images/}}

\title[Managing Crowded Museums]{
       Managing Crowded Museums: Visitors Flow Measurement, Analysis, Modeling, and Optimization}

\author[P. Centorrino, A. Corbetta, E. Cristiani, and E. Onofri]{}


\begin{document}
\maketitle

\centerline{\scshape Pietro Centorrino}
\medskip
{\footnotesize
	\centerline{Department of Physics, Sapienza Universit\`a di Roma, Rome, Italy}
	\centerline{\texttt{pietro.centorrino@gmail.com}}
}
\bigskip
\centerline{\scshape Alessandro Corbetta}
\medskip
{\footnotesize
	\centerline{Department of Applied Physics, Eindhoven University of Technology, Eindhoven, The Netherlands}
	\centerline{\texttt{a.corbetta@tue.nl}}
}
\bigskip
\centerline{\scshape Emiliano Cristiani}
\medskip
{\footnotesize
	\centerline{Istituto per le Applicazioni del Calcolo, Consiglio Nazionale delle Ricerche, Rome, Italy}
	\centerline{\texttt{e.cristiani@iac.cnr.it}}
}
\bigskip
\centerline{\scshape Elia Onofri}
\medskip
{\footnotesize
	\centerline{Istituto per le Applicazioni del Calcolo, Consiglio Nazionale delle Ricerche, Rome, Italy}
	\centerline{\texttt{eonofri@uniroma3.it}}
}

\bigskip

\begin{abstract}
We present an all-around study of the visitors flow in crowded museums: a combination of Lagrangian field measurements and statistical analyses enable us to create stochastic digital-twins of the guest dynamics, unlocking comfort- and safety-driven optimizations.
Our case study is the Galleria Borghese museum in Rome (Italy), in which we performed a real-life data acquisition campaign. 

We specifically employ a Lagrangian IoT-based visitor tracking system based on Raspberry Pi receivers, displaced in fixed positions throughout the museum rooms, and on portable Bluetooth Low Energy beacons handed over to the visitors.
Thanks to two algorithms: a sliding window-based statistical analysis and an MLP neural network, we filter the beacons RSSI and accurately reconstruct visitor trajectories at room-scale. Via a clustering analysis, hinged on an original Wasserstein-like trajectory-space metric, we analyze the visitors paths to get behavioral insights, including the most common flow patterns.
On these bases, we build the transition matrix describing, in probability, the room-scale visitor flows. Such a matrix is the cornerstone of a stochastic model capable of generating visitor trajectories \textit{in silico}.
We conclude by employing the simulator to enhance the museum fruition while respecting numerous logistic and safety constraints. This is possible thanks to optimized ticketing and new entrance/exit management.

\end{abstract}

\tableofcontents

\section{Introduction}
The analysis of the behavior of museum visitors has a long-standing tradition~\cite{robinson1928book} and grows daily in importance as tourist flows increase and digital technologies get  ubiquitous~\cite{falk2016book, yalowitz2009}.
The outstanding issue of visitors management demands for multidisciplinary skills connected to, among others, psychology, computer science, statistics, physics of complex systems as well as modeling and optimization theory. 

Museums curators are expected to achieve three complex and seemingly
contradictory objectives: increasing the visitors number, enhancing the experience quality, and preserving the artworks~\cite{yoshimura2014}. 
Accurately measuring and analyzing the visitors trajectories is an essential component towards these objectives and, specifically, when aiming at an efficient organization of the exhibits \cite{bourdeau2001, trondle2014}, the determination of adequate ticketing strategies, and also to verify if visitors experience complies with managers' intents~\cite{tzortzi2014}. 

\medskip
A complete workflow enabling the full control of visitors in a museum consists of several challenging steps, that we here summarize.

\begin{description}[labelindent=\parindent,leftmargin=0cm]%
\item[Visitors tracking] the first goal is to understand the behavior of visitors in terms of paths followed in the museum. Not all museums have predefined paths and sometimes more than one choice is possible \cite{lanir2017}. Moreover, in large museums, it is rare that visitors see the whole exhibition \cite{piccialli2020b}. 
A number of technologies exist for indoor tracking that is characterized by a trade-off between  deployment complexity, invasiveness and accuracy. 
Radio-based approaches, as considered in this work, enable room-level positioning accuracy: visitors trajectories are rendered into sequences of visited rooms and related permanence times. 
At the price of more invasive and complex deployments, sometimes impossible in the context of cultural heritage, centimeter-level individual positioning can be also accomplished, e.g.\ via distributed grids of 3D scanners or video-cameras.
\\
Besides, psychological and sociological variables can be observed on side of paths, such as heart rate, skin conductance, emotional and aesthetic evaluations of specific artworks~\cite{kirchberg2015, trondle2014, kirchberg2014}, interactions with groupmates, degree of attention, boredom or fatigue. \\
Automatic systems could be complemented with manual activities, like paper-and-pencil annotations and questionnaires \cite{klein1993, mokatren2019, trondle2014}.
From questionnaires, one can estimate demographic related and museum visit related features \cite{mokatren2019} like age, gender, educational level, number of visits per year to museums, etc. 
After the visit, one can measure the degree of satisfaction, the relationship between perceived and real time spent in the museum \cite{bollo2005}, etc.
\item[Behavior understanding]
a number of variables can be estimated from visitors trajectories: busy hours, movement patterns, length of visits, permanence times in each room, number of stops.
Two indicators are generally considered to quantify the importance of a specific exhibit, the \emph{attraction power} (relative amount of people who has stopped in front of an artwork during their visit) and the \emph{holding power} (average time spent in front of an artwork) \cite{lanir2017}.
\\
Clustering and AI-based algorithms can be used for inferring, from the whole trajectories dataset, the \emph{typical paths} or, equivalently, the \emph{typical individual behaviors} inside the museum.
Another interesting question regards the predictability of visitors behaviors \cite{cuomo2016, kuflik2012, martella2017}: can a person who starts visiting the museum in a certain manner be immediately labeled as a visitor of a certain type? \\
\emph{Social behavior} can be observed too. For example, one can wonder, e.g., if people belonging to the same group follow the same path or they split, or whether individuals are attracted or repelled by crowding.
\item[Museums digital twin]
once statistics about visitors trajectories and behavior are available, it is possible to create an algorithm capable of generating real-like visits paths in the museum \cite{balzotti2018}.
This is done by reproducing the movements of people from one room to another, duly determining their transition probability.
Moreover, herding behavior in social groups or the response to congestion and fatigue could be taken into account. A digital twin is able to reproduce virtual visitors moving in the museum with a realistic behavior, possibly in new (i.e.\ unexperienced) conditions. It is also possible to forecast the visitors flow from some initial conditions, like, e.g.\ the visitor inflow at a given time.
\item[Visitors flow optimization] 
in order to use the museum digital twin as managing tool, curators and organizers have to identify relevant \emph{control variables} and \emph{objectives}: regarding the former, one can, e.g., regulate the entrance flows, limit the maximum occupancy of selected rooms, increase the number of entrances or exits, set a maximal duration of the visit. The ticket price can obviously be controlled too. \\
Regarding objectives instead, one can aim at maximizing the number of visitors, the pleasantness of the visit, the amount of information conveyed, or keeping the environmental parameters (e.g., temperature and humidity) in a given range, for best conservation of the collection.
\\
Once this is done, a museum digital twin can be profitably used to simulate different scenarios, aiming at matching the objectives while varying  the control variables. Here optimization algorithms like gradient-based or PSO methods can be used to automatize the search for a solution.
\end{description}

\subsection{Relevant literature}\label{sec:refs}
The first step (Visitors tracking) is the one that has received the most attention in the literature, as it relates to pedestrian dynamics in general, i.e.\ beyond the museum context.

Focusing on (indoor) tracking systems, all kind of technologies have been exploited, such as
RFID \cite{lanir2017},
Wi-Fi \cite{georgievska2019, hong2018},
Bluetooth \cite{casolla2020, choi2017, versichele2012a, martella2017, oosterlinck2017, piccialli2020, piccialli2019a, pierdicca2019, piccialli2019b, versichele2012b, yoshimura2012, yoshimura2014},
video cameras \cite{lovreglio2018},
3D scanners \cite{corbetta2018, seer2014}.
An exhaustive review of these methods is out of the scope of the paper; we refer the interested reader to the papers \cite{gu2009, oosterlinck2017} for more references.

Different technologies require different degree of visitors involvement. 
For example, video cameras, 3D scanners, Wi-Fi or Bluetooth mass scans require no collaboration, while Bluetooth-based apps and RFID tags usually require some degree of visitors interaction.
Measuring personal data like heart rate or skin conductance requires instead total involvement \cite{kirchberg2015, kirchberg2014}.
Moreover, convincing people to participate in an experiment, for example by downloading and installing a smartphone app, can be difficult and time-consuming \cite{pierdicca2019}.
Sometimes free tickets could yield a good incentive \cite{zancanaro2007}.

\medskip

The second step (Behavior understanding) has also been investigated in great detail in connection with museums. Regarding \emph{individual behavior}, the predominant idea is to classify visitors into four categories based on the way they interact with the artworks:
`Ants' (tend to follow a specific path and observe extensively almost all the exhibits);
`Butterflies' (do not follow a specific path but are guided by the physical orientation of the exhibits;  stop frequently to acquire more information);
`Fish' (most of the time move around in the center of the room and usually avoid looking at exhibits details); 
`Grasshoppers' (seem to have a specific preference for some pre-selected exhibits and focus their time on them, while tending to ignore the rest),
see, e.g.,~\cite{kuflik2012} or~\cite{veron1989book} for the origin of this taxonomy.

Regarding \emph{social behavior} instead, the idea is to label visitors in six categories based on how they interact with group mates:
`Doves' (interested in other visitors while ignoring the environment);
`Meerkats' (stand side by side, expressing great interest in the exhibits);
`Parrots' (share their attention between exhibits and group members);
`Geese' (advance together, however one visitor appears in the lead);
`Lone wolves' (enter the museum together and then separate); 
`Penguins' (cross the space together while ignoring the exhibits),
see, e.g., \cite{dim2014, lanir2017}.

Clustering techniques (e.g., $k$-means, hierarchical clustering, sequence alignment) have been used to assign every visitor trajectory (spanning from room-scale to continuum) to one of the groups described above, or to some given typical movement patterns \cite{casolla2020, dim2014, versichele2012a, kuflik2012, mokatren2019, martella2017, oosterlinck2017, piccialli2019a, piccialli2019b, yoshimura2019, zancanaro2007}.
This enables one to quantify the percentage of visitors belonging to each group.
Note that typically the number of clusters is assigned \emph{a priori} and this can be an important limitation.
One crucial point for cluster investigation is the definition of a suitable \emph{metric}, to measure the distance between trajectories, and aggregate (cluster) trajectories close to each other. 
Examples of such metrics devised at room-scale can be found in \cite{casolla2020, martella2017, piccialli2019a}.
In particular, \cite{casolla2020, piccialli2019a} propose a combination of well-known metrics defined in the space of characters strings (as trajectories can be suitably represented as sequences of characters), which is further corrected to take into account the differences in time of permanence in each room.

Regarding trajectory comparisons, let us mention also other two papers: \cite{yoshimura2019} compares measured trajectories with those coming from a random walk simulator in order to understand which kind of visitors exhibits stronger patterns.
The work~\cite{lanir2013} compares trajectories of visitors with and without audio-guides in order to measure the impact of the transmitted information.

\medskip

The third and fourth steps are also related to the rich pedestrian flow modeling literature:
if one considers the museum as a continuous space as in \cite{liakou2019}, one can refer to differential (agent-based, kinetic, fluid-dynamic) or nondifferential (discrete choice, cellular automata) models.
See, e.g., \cite{bellomo2011, cristiani2014book, duives2013, dong2020, eftimie2018, haghani2020, martinez-gil2017} for some reviews, books' chapters and books about this topic.

If one instead considers the museum as a graph -- where the nodes represent the rooms of the museum and the edges represent connections among rooms -- one can refer to some classical tools like transition matrices and deterministic/stochastic Markov chains with/without memory \cite{hong2018, piccialli2020, lees2016} in order to simulate a room-level walk in the museum (i.e.\ a trajectory on the graph).

Although many mathematical tools are available, examples of actual museums digital twins developed with the aim of reproducing, understanding and optimizing visitors behavior are largely missing.
This fact holds despite the fact that the path followed by visitors is evidently conditioned by the design of the exhibition galleries \cite{bourdeau2001, trondle2014}.
An interesting attempt can be found in \cite{guler2009thesis, guler2016book}: the author describes a museum simulator and uses it to show that changes in the layout design of an exhibition result in different visitor circulation patterns. 
Unfortunately, that simulator can be hardly used in a museum with a very high density of artworks exposed, since it requires a complex calibration of many artwork-scale parameters which usually show a high variance between visitors.   
See also \cite{balzotti2018} for a rudimentary simulator on graph and \cite{pluchino2014} for a simulator developed under the NetLogo software environment.

\subsection{Paper contributions}
In this paper, we perform an all-around investigation which includes contributions to all the four steps described above. 
Covering the whole process allows us to reach an unprecedented level of understanding and control of the museum, which unleashes the capability of improving deep modifications to the ticketing strategy as well as to the museum access management. 
Our results are based on real visitors data acquired in the Galleria Borghese museum (Rome, Italy).
In more details, the research unfolds along the following lines:
\begin{enumerate}[labelindent=0cm,wide]
\item
We describe a cheap and easily reproducible data collection system, hinged on an IoT-based room-scale Lagrangian tracking system of the museum visitors.
Each visitor is provided with a portable Bluetooth Low Energy (BLE) beacon, whose signal is received by antennas (realized by means of common Raspberry Pi's)  displaced in fixed positions within the museum rooms. We employ this system for an extended data collection campaign which provides the high statistics measurements employed in this work.
\item
We employ and filter the Received Signal Strength Indicator (RSSI) of the beacons to reconstruct individual visitors trajectories.
Due to the restricted space and the numerous architectural and historical constraints, each beacon is often captured by multiple antennas at the same time.
Accurately reconstructing the trajectories in this settings defines a challenge \textit{per se}.
We propose a new machine learning approach which outperforms standard sliding window processing, especially, when it comes to estimating the correct time of permanence in rooms.
\item
In order to get insights about visitors behavior, including the most common movement patterns, we analyze trajectories via statistical and clustering techniques.
Inspired by the Wasserstein distance\footnote{The Wasserstein distance was first introduced by Kantorovich in 1942 and then rediscovered many times. Nowadays, it is also known as $Lip'$-norm, earth mover's distance, $\bar d$-metric, Mallows distance. An important characterization is also given by the Kantorovich--Rubinstein duality theorem.}, we introduce a new \emph{ad hoc} trajectory clustering metric, which respects the geometrical properties of the museum. Our metric, in fact, builds upon the physical distance among rooms. 
By using a hierarchical cluster analysis, only based on such  metric (and with no \emph{a priori} hypothesis on the number of clusters nor on their size), we can unveil automatically hard-to-see movement patterns that go well beyond the standard animal-inspired classification (see Section \ref{sec:refs}).  
As a by-product, we can also identify anomalous behaviors.
\item
We employ statistical tools to build a probability transition matrix among museum rooms, which provides us with building blocks for a model capable of simulating \textit{in silico} the museum visits. 
In particular, this enables us to forecast the path of visitors entering the museum from any room.
Unlike the simulator presented in~\cite{guler2009thesis, guler2016book}, our simulator leverages on the measured permanence time in each room on side of the probability of transition from one room to any other. This results in a tool easier to calibrate. 
\item
Finally, we employ the simulator to significantly increase the efficiency of the ticketing strategy and entrance/exit management. Our results suggest a way to enhance the museum fruition from both visitors and curators points of view, while keeping the numerous constraints within the limits.
\end{enumerate}

\subsection{Paper organization}
We present our methods and original contributions alongside our field activity at Galleria Borghese museum, our case study. 
In Section \ref{sec:casestudy} we introduce Galleria Borghese and outline its floor plan and the current ticketing strategy.
In Section \ref{sec:trackingsystem} we describe our tracking system and the dataset we collected in the museum.
In Section \ref{sec:trajectoryreconstruction} we discuss our trajectory reconstruction methods.
In Section \ref{sec:trajectoryanalysisandclustering} we analyze the trajectories collected, fit their statistical observable with known distributions, and introduce our clustering approach.
In Section \ref{sec:model} we introduce the model which allows us to create a complete digital twin of the museum and simulate \emph{in silico} the visitors flow.
In Section \ref{sec:optimization} we employ the model to find optimal  strategies for ticketing and museum management.
The discussion in Section \ref{sec:conclusions} closes the paper.

\section{Case study: the Galleria Borghese in Rome}\label{sec:casestudy}
\begin{figure}[t]
	\begin{center}
		\textbf{a.} \includegraphics[width=0.438\linewidth]{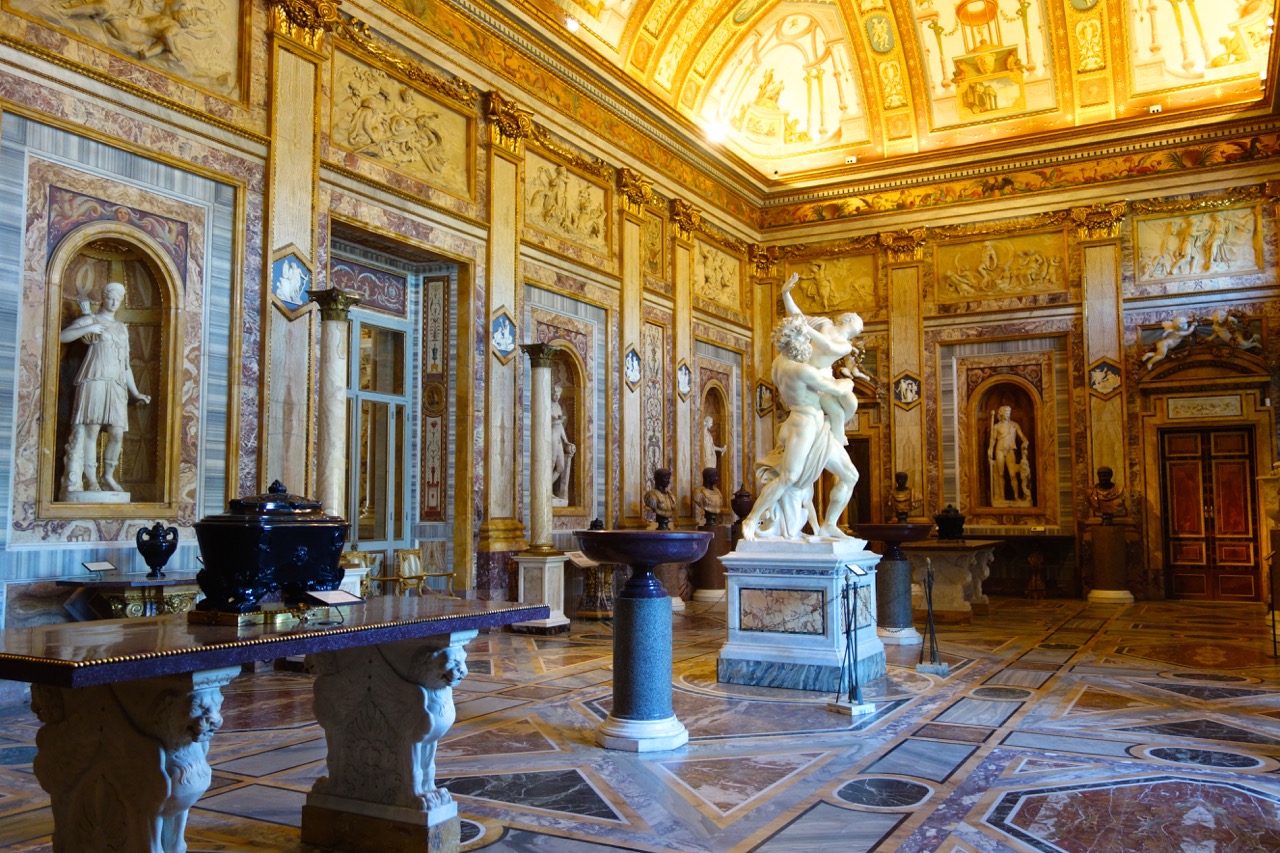}
		\textbf{b.} \includegraphics[width=0.47\linewidth]{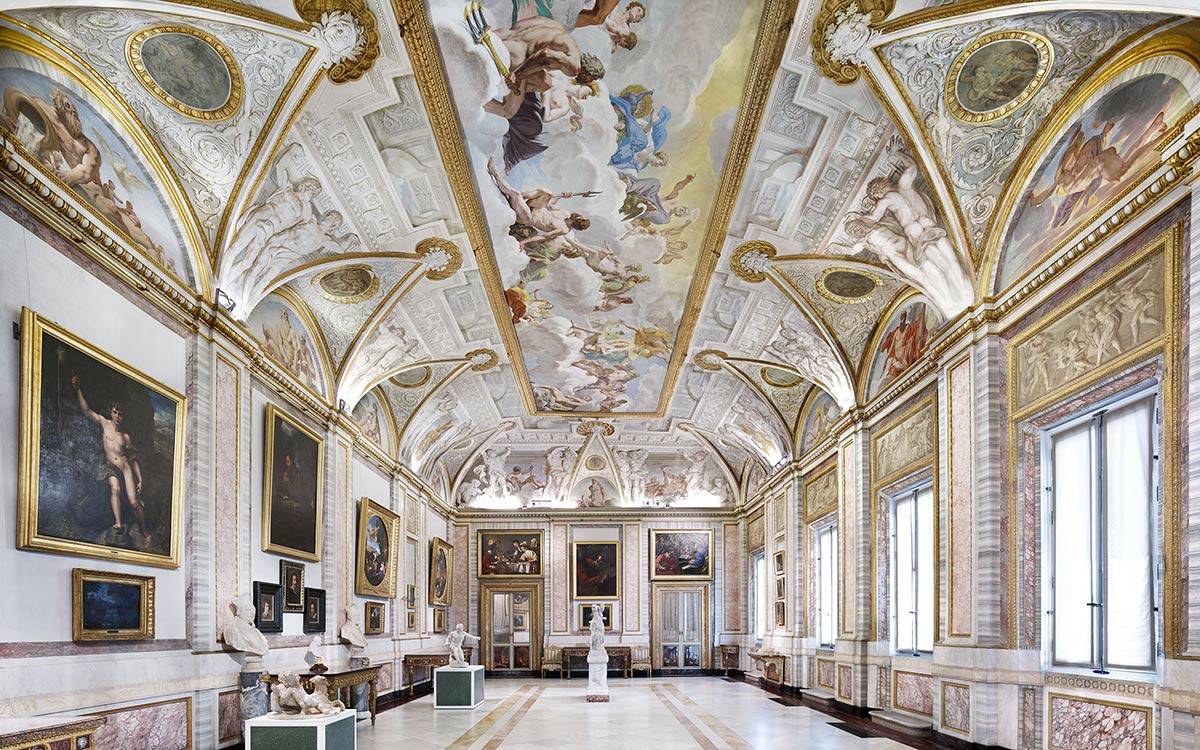}
		\caption{The two largest rooms in Galleria Borghese. \textbf{a.} \emph{Ratto di Proserpina} located on the first floor. \textbf{b.} Main area on the second floor, part of the Pinacoteque. The map of the museum and the room names will be shown later in Figure \ref{fig:GBmap} and Table \ref{tab:matchroomsantennas}.}
		\label{fig:splendoreGB}
	\end{center}
\end{figure}

The world-renowned Galleria Borghe\-se museum (Rome, Italy), is a relatively small, two-floor museum with 3 entrances and 21 exhibition areas. 
Its sculptures and paintings attract visitors from all over the world, see Figure \ref{fig:splendoreGB}.
On the main floor, the exhibition area is circular, while on the second floor (Pinacoteque) it is U-shaped. 
Rooms are numbered but no obligatory exhibition path is assigned, so many people do not visit the rooms in their natural order.
Moreover, the density of exhibits is so high that people often come back to already visited rooms multiple times to admire artworks missed during the previous passages.
Congestion is frequent in some rooms, like the one which hosts Caravaggio's paintings.
Audio-guides are available on-demand and guided tours are allowed but subject to quota (both in number and size).

To cope with the many historical, artistic and architectural constraints, museum curators established to schedule the visits: tickets must be booked in advance and give access to the museum for a slot of 2h. 
Five slots per day are granted. The maximum number of visitors allowed in each slot is 360. Additionally, 30 tickets, called ``last-minute'', are sold 30 minutes after the beginning of each time slot.
People can also decide which floor to start the visit from, within some limits.
At the end of each time slot, people are invited to leave, and the museum empties completely.
Let us also note that many visitors enter without their smartphone since they must leave their personal bags in the wardrobe. 

It is plain that Galleria Borghese has specific peculiarities which make it rather unique in the world. This means that not all aspects of the present study are directly applicable to other museums. Let us mention, in particular, the entrance system with quota and the fact that visitors often return to already visited rooms. 
On the other hand, other aspects are easily generalizable. Mathematical and numerical methods presented here can be used whenever one has to track people moving in a built environment through a nonpredefined sequence of rooms. In addition, control and optimization technique are suitable whenever the flow of people can be controlled in some way, e.g.\ changing entrance doors and/or changing the visiting path dynamically.

\section{Data collection: IoT visitors tracking system}\label{sec:trackingsystem}

As many cultural heritage sites worldwide, Galleria Borghese is covered by frescoes and paintings, which heavily limit the possibility of displacing  (electrical) devices. To cope with these historical and architectural constraints, we developed a noninvasive radio-based IoT measurement solution delivering room-level visitors trajectories.

\begin{figure}[t]
	\centering
	\textbf{a.} \includegraphics[width=0.36\linewidth]{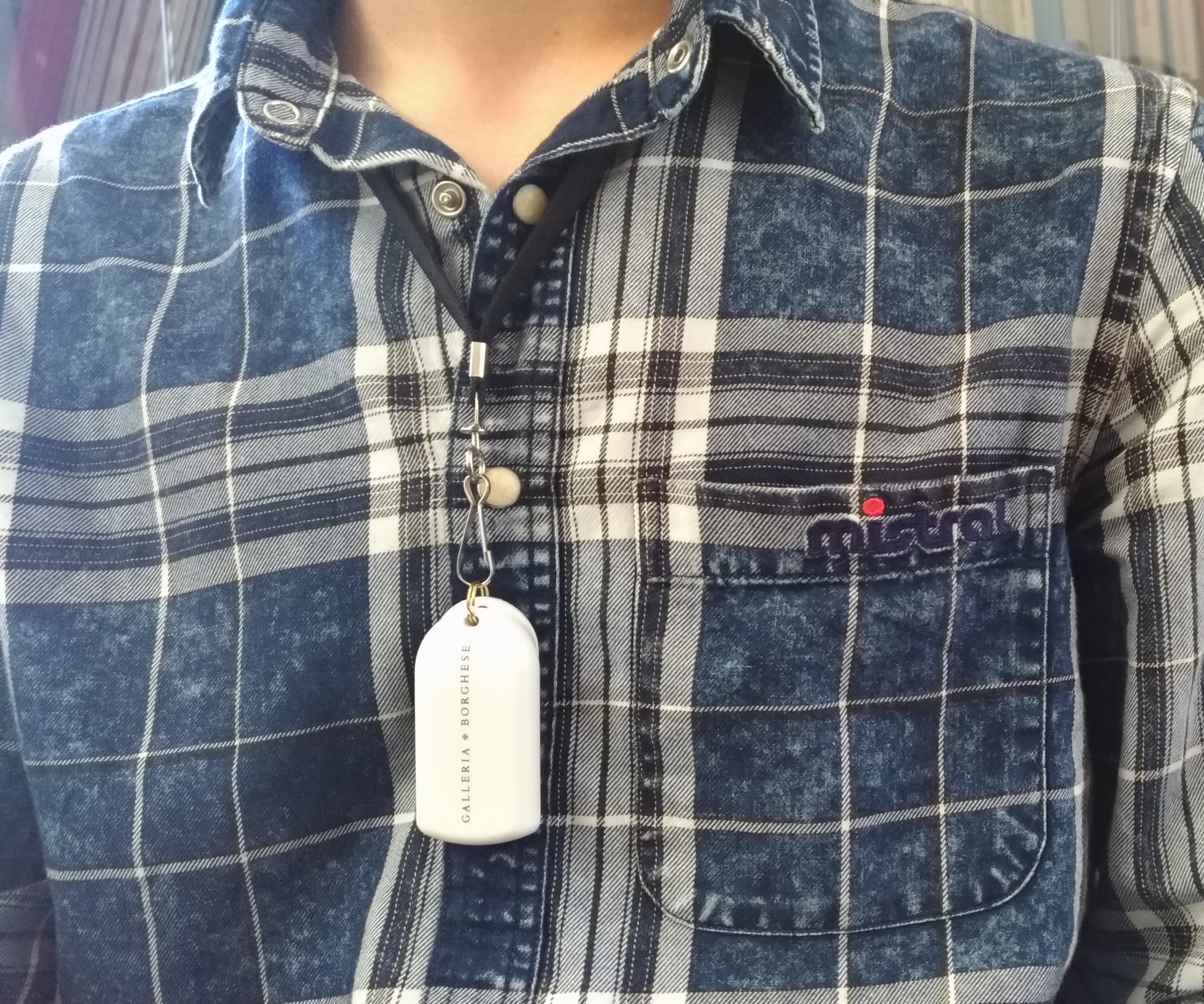}\hfill
	\textbf{b.} \includegraphics[width=0.48\linewidth]{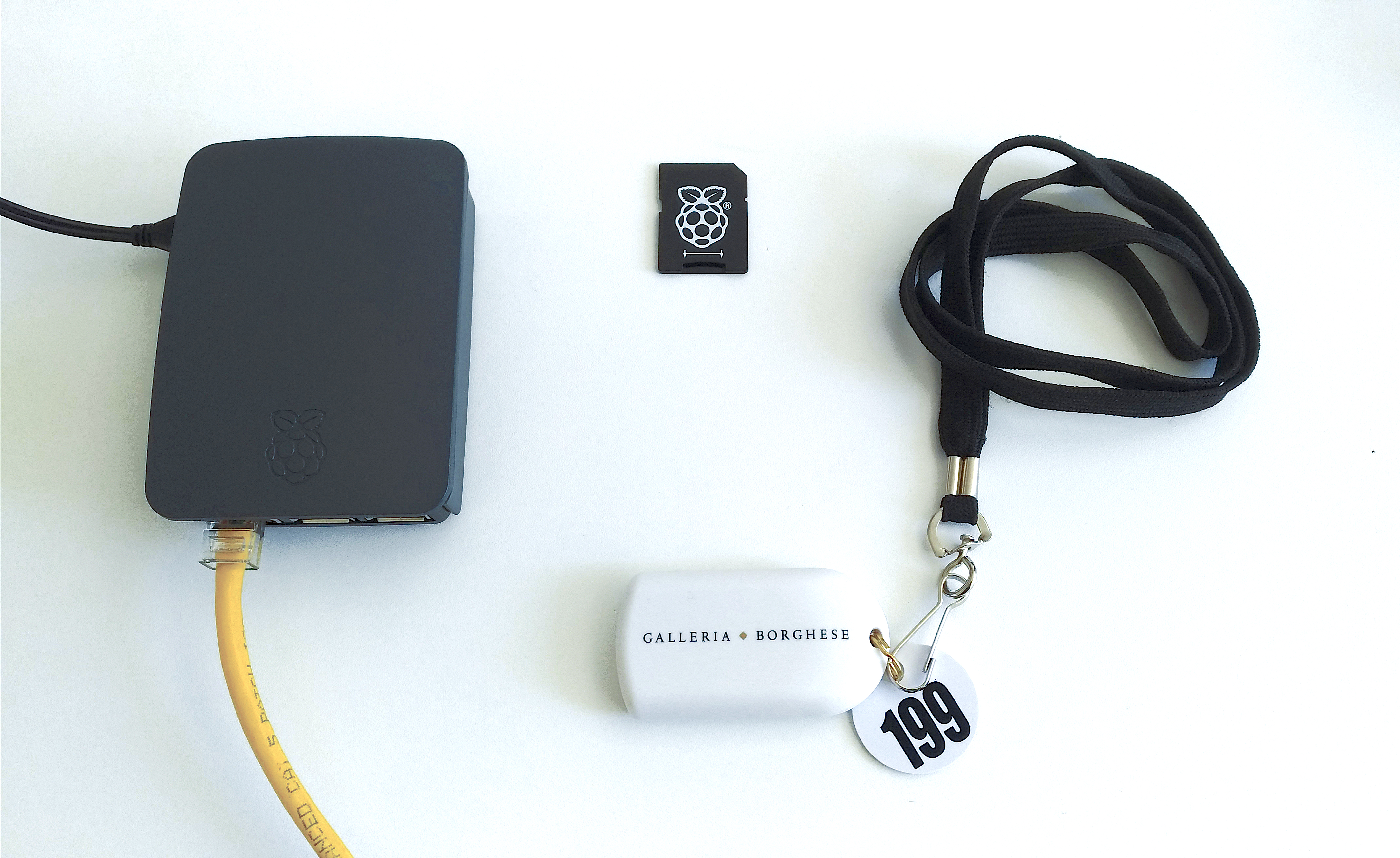}
	\caption{\textbf{a.} Sample visitor wearing the BLE beacon. \textbf{b.} Raspberry Pi used as Bluetooth antenna to receive beacon signals and measure their RSSI.}
	\label{fig:trackingsystem}
\end{figure}

Figure \ref{fig:trackingsystem} shows the main components of the tracking system, which consists of:
\begin{description}
	\item[Transmitters]
	we gave a small BLE beacon to each visitor to track who was briefed about the experiment, see Figure \ref{fig:trackingsystem}\textbf a.
	The beacon transmitted a signal at $+4$dB with iBeacon standard encoding \cite{newman2014}, which 
	carries a unique identifier (UUID).
	\item[Fleet of receiving antennas]
	we employed RaspberryPi 3B+ (RPi) as receivers. A RPi is a single-board computer with embedded Bluetooth and Wi-Fi modules, see Figure \ref{fig:trackingsystem}\textbf b.
	RPi's were located along the museum in fixed positions, see Figure \ref{fig:GBmap}.
	A Python code running on the RPi's was used to scan continuously the surrounding area listening for beacons signals. Each signal is stored as a tuple containing the beacon's identity, the RSSI and the timestamp of reception.
	Every 5s a data packet was created (with only one occurrence of each beacon detected), it got signed with the RPi identity and posted on a central server via an internet connection.
	\item[Central server]
	the server received data packets from all RPi's and stored them in a SQL database along with the reception timestamp. Such couple of timestamps allows us to quantify the duration of the whole process.
\end{description}

The data presented in this paper come from a measurement campaign lasted between June and August 2019.
The central SQL server received $1,308,617$ records corresponding to $900$ visitors trajectories surveyed during $13$ 2h-long visit slots.
The percentage of tracked visitors w.r.t.\ the total number was about 1:5. %
As it usually happens, the vast majority of visitors came in groups (family, friends, guided tours, etc.).
In this case, apart from a few exceptions, we tracked only one member of each group, thus losing the ability to detect the interactions within social groups. 	

We have also tested that collecting dozens of beacons at the same time within a small physical space does not impact on the reliability of the system. 

\begin{figure}[t]
	\begin{center}
        \includegraphics[width=0.75\linewidth]{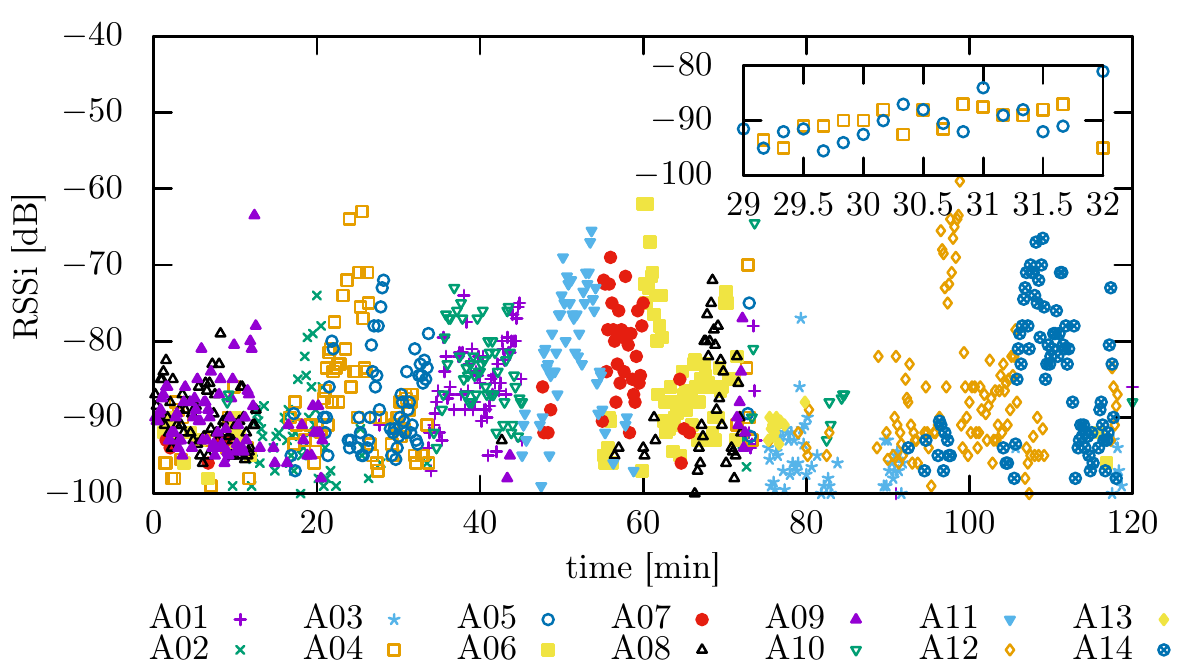}
		\caption{Typical raw RSSI throughout a 120 minute visit as recorded by the 14 RPi's receiving antennas. 
		(Re-sampled every $\Delta t=10\,$s; RPi's antennas are distinguished by markers).
		Inset: between minute 29 and 32 the visitor is detected by both RPi4 and RPi5, and the maximal RSSI strongly oscillates between the two. As a consequence, the signal strength is insufficient to associate unambiguously a visitor to a location.
        }
		\label{F:RawBeacon}
	\end{center}
\end{figure}

Figure \ref{F:RawBeacon} shows the history of a single beacon's RSSI (i.e., a single visitor) during a visit.
The analysis of the raw data immediately confirms that RSSI signal suffers from high fluctuations (see also \cite{beder2012}). This means that a beacon fixed in the middle of a room is not received with a constant RSSI, and the RSSI of two equidistant beacons might not be the same.
In addition, we encountered other two important difficulties:
\begin{enumerate}
	\item 
	A single beacon can be detected by multiple antennas at the same time. RSSI is used to resolve the ambiguity but high fluctuations make such task rather hard.
	\item Some areas of the museum could not be covered at all (e.g., staircase between the two floors, due to lack of electrical outlets). 
\end{enumerate}

Finally, let us also mention the possibility -- which we consider very rare -- that visitors wearing beacons could be influenced by the fact that they feel tracked, cf.\ \cite{trondle2014, yoshimura2014}.

In the next section, we describe how the raw RSSI data from the transmitters is processed to estimate individual trajectories. 

\section{Trajectory reconstruction and filtering}\label{sec:trajectoryreconstruction}
We use the raw data collected by the tracking systems to reconstruct the sequence of visited rooms and the time of permanence in each room.

First of all, we achieve uniform temporal sampling of the signals through a re-sampling in bins of fixed time length $\Delta t$ = 10s (cf.\ Figure \ref{F:RawBeacon}). $\Delta t$ is to be tuned according to both the resolution needed and the signal granularity.
We employed a -120 dB threshold for those antennas not detecting the beacon.
The output of this procedure is a $A\times T$ matrix $\RSSI$ for each beacon, where $A$ is the number of antennas and $T$ is the number of time bins (duration of the visit divided by $\Delta t$). In other words, for a given beacon, the element $\RSSI_{a,t}$ is the RSSI of the signal received by $a$-th antenna in the $t$-th time bin.

\begin{figure}[t]
	\begin{center}
		\def\svgwidth{0.75\linewidth}
			\input{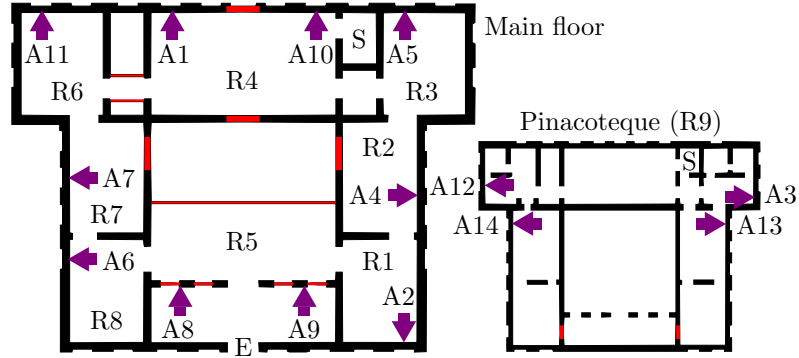}
			\caption{
				Floor plan of Galleria Borghese.
				Rooms (R) and receiving antennas (A) are reported (cf.\ Table~\ref{tab:matchroomsantennas} for room-antennas relations).
				Red lines represent closed passages/doors.
				Visitors admittance happens both at the main entrance on the first floor (E) and at the stairs (S) on either floor.
			}
		\label{fig:GBmap}
	\end{center}
\end{figure}

In order to simplify our room-scale tracking, we have merged the 21 exhibition areas of the museum into $R=9$ (radio) rooms in which we deploy our $A=14$ receiving antennas (see Figure \ref{fig:GBmap} and Table \ref{tab:matchroomsantennas} for antennas positions and antenna-room assignments). 
We remark that signal readings in a given room do not imply that the emitting beacon is located in the same room.

\begin{figure}[t]
	\centering
	\includegraphics[width=0.985\linewidth]{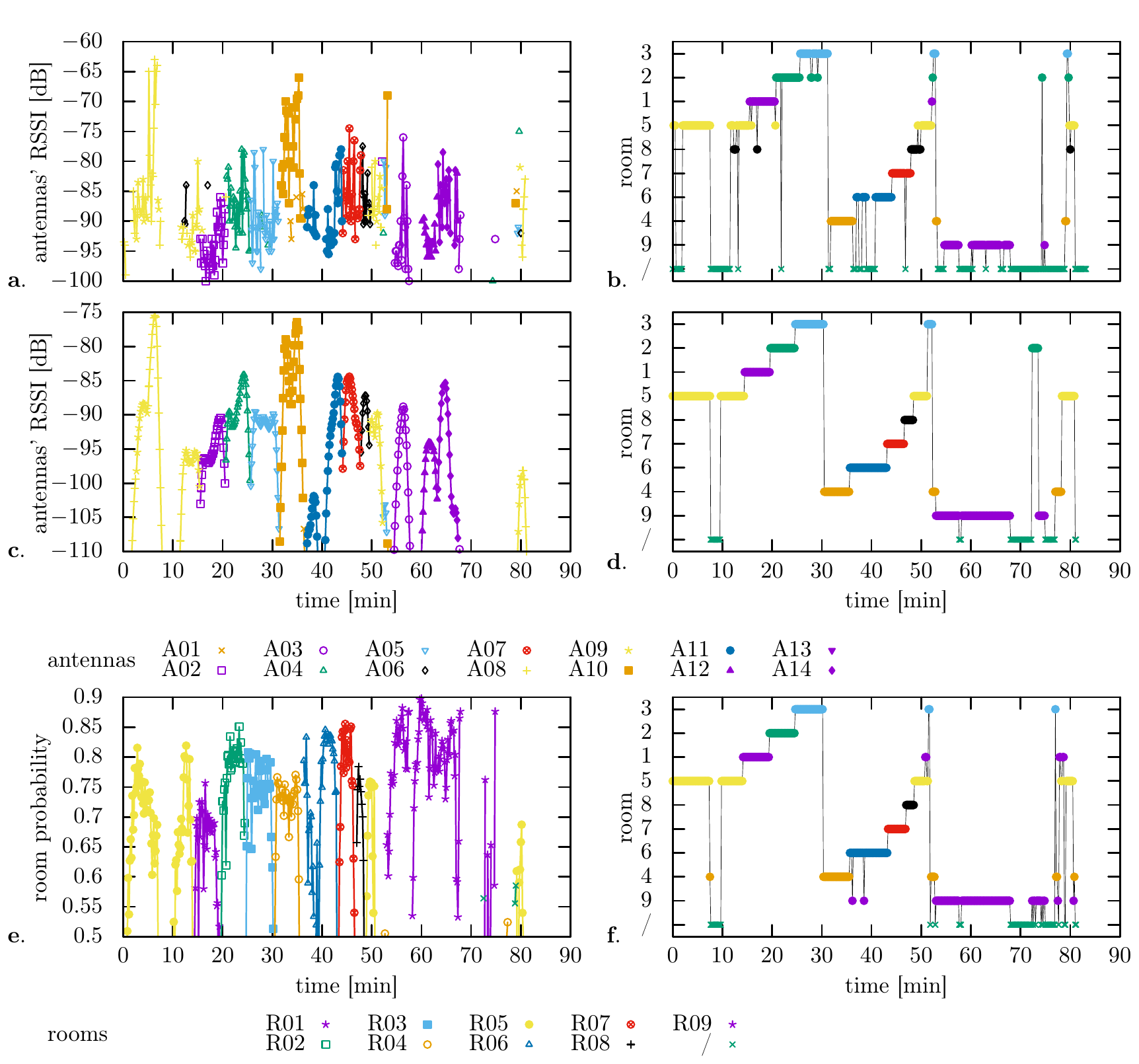}
	\caption{
		A sample beacon RSSI elaborated by argmax (\textbf{a.}\ \& \textbf{b.}), sliding window (\textbf{c.}\ \& \textbf{d.}), and machine learning (\textbf{e.}\ \& \textbf{f.}) approaches. 
		The left column reports the max of RSSI for the argmax and sliding window approaches (antennas located in the same room are labeled with the same color but different markers), and the maximum among the rooms probabilities for the machine learning approach. 
		The right column reports the corresponding reconstructed trajectories as sequence of rooms.
		Not-detected statuses are marked by green crosses ({\color{ForestGreen}$\times$}). 
	}
	\label{F:TrElaboration}
\end{figure}

\begin{table}[t]
    \begin{center}
        \begin{tabular}{ l l l }
            \hline
            Number &  Room  nickname    &   Antenna    \\
            \hline\hline
            R1  &   Paolina             &   A2         \\
            R2  &   David               &   A4         \\
            R3  &   Apollo e Dafne      &   A5         \\
            R4  &   Ratto di Proserpina &   A1, A10    \\
            R5  &   Portico         &   A8, A9    \\
            R6  &   Enea e Anchise      &   A11         \\
            R7  &   Satiro su delfino   &   A7         \\
            R8  &   Caravaggio          &   A6         \\
            R9  &   Pinacoteque         &   A3, A12, A13, A14  \\
            \hline
        \end{tabular}
    \end{center}
    \caption{Match among rooms (R) and RPi antennas (A) in Galleria Borghese museum.}
    \label{tab:matchroomsantennas}
\end{table}
The most natural way to reconstruct visitor trajectories is to compute the argmax of the RSSI history of each beacon:
at each time bin one retains the antenna that receives the highest RSSI. 
Finally, one defines the current visitor location as the room associated to the antenna.
The result of this procedure is shown in Figure~\ref{F:TrElaboration}\textbf{ab}.
Unfortunately, when a visitor is at approximately equidistant from two or more RPi's, the maximal RSSI quickly bounces back and forth among the antennas. In signal terms, the visitor appears to perform extremely rapid and unrealistic room changes.

Building upon~\cite{centorrino2019}, we consider two data refinement methods: the first one relying on a sliding window approach and the second one based on a neural network.

\subsection{Sliding window approach}\label{ssec:SW}
The first method aims at smoothing the noise in the RSSI data by applying a low-pass filter, implemented as a \emph{weighted moving average}, and a \emph{normalization}.
RSSI's gathered close in time should have close values; besides, the closer the bins, the higher is the correlation.  

In particular, we convolve the RSSI signals gathered by each antenna (i.e.\ the matrix rows $\RSSI_{\cdot, t}$) with a (symmetric triangular) kernel with size $2\delta + 1$ and weights $w_{0},w_{1},\ldots,w_{2\delta}$. 
In formulas, this approach generates a new matrix $\tilde \RSSI$ defined as
\begin{equation}\label{E:scalarTildeR}
	\tilde \RSSI_{a, t} =
		\sum_{d = t-\delta}^{t+\delta}
		\RSSI_{a, d} \cdot w_{\delta - t + d},
	\quad
	0 \leq a < A, \quad
	\delta \leq t < T - \delta \ .
\end{equation}

Secondly, a normalization is applied across the signals acquired by the different antennas in order to make them comparable.
This produces a third matrix
$\bar \RSSI$ as
\begin{equation}
	\bar \RSSI_{a, t} = \frac{\tilde \RSSI_{a, t} - \mu_t}{\sigma_t}
	, \qquad
	0 \leq a < A
	, \quad
	\delta \leq t < T - \delta\ ,
\end{equation}
where $\mu_t$ and $\sigma_t$ are respectively the mean and the standard deviation of $\bar \RSSI$ by time bin  (i.e.\ by column, thus $\mu_t = \mu(\bar \RSSI_{\cdot,t})$, $\sigma_t = \sigma(\bar \RSSI_{\cdot,t})$).
Figure~\ref{F:TrElaboration}\textbf{cd} shows the result of this procedure.

\subsection{Machine learning approach}\label{MLapproach}
To improve performances of the sliding window method, we propose a trajectory reconstruction approach based on neural networks.
At any time bin $t$, we cast the localization of a visitor in one among the $R$ rooms as a classification problem. Our neural network processes the $\RSSI$ matrix (in time windows) and returns the probability vector whose $r$-th component is the probability that the visitor is located in the room $r$. 

\subsubsection{Building the neural network}

We consider a neural network made of $L + 1$ layers, with $L=2$. The data is injected in the first layer and flows ``forward'' in the network through the hidden layer to the output layer.
Each layer $\ell$ is built out of a different number $n_\ell$ of nodes
$a^{(\ell)}$, that represent the calculus units of the network, or artificial neurons,
where $a_j^{(\ell)}, 1 \leq j \leq n_\ell$ represents the $j$-th neuron of layer $\ell$.

\begin{figure}[t]
	\begin{center}
		\scalebox{0.7}{
			\input{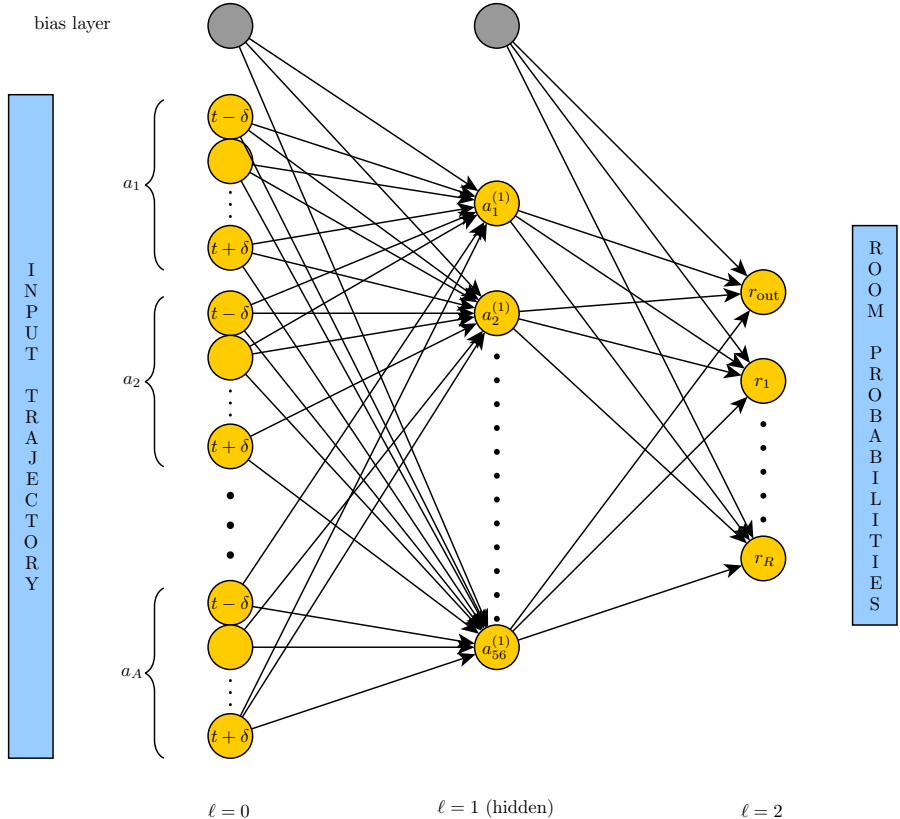}
        }
		\caption{
			The three layers ($L=2$) neural network employed to process the trajectories collected in Galleria Borghese. For each time bin, the neural network predicts the visitor position by considering the RSSI within the previous and the next minute.
			The input layer is made of $(2\delta + 1) \times A = 182$ neurons, where $\delta = 6$ is the semi-amplitude (one-minute long) of the sliding time interval for each of the $A=14$ antennas. 
			The output layer is made of $R=10$ neurons, one for each room of the museum plus the ``out'' condition.
			The single hidden layer is composed by $56 = 14 \times 4$ neurons, as a trade-off between the input and the output layer sizes.
		}
		\label{F:NeuralNetwork}
	\end{center}
\end{figure}

The specific network that we employ, also known as Multi Layer Perceptron (MLP), is built as a complete weighted directed graph between the nodes within  layer $0 \leq \ell < L$ and the nodes of the next layer $\ell+1$ (cf.\ Figure \ref{F:NeuralNetwork}).
A spare node with fixed value $1$ (\emph{bias}) and index $0$ is added to each layer but the last, that is $a^{(\ell)}_0 = 1, 0 \leq \ell < L$.
We denote by $\Theta_{s, d}^{(\ell)}$ the weight of the edge directed from the $s$-th node of the $\ell$-th layer  to the $d$-th node of the $(\ell+1)$-th layer.

\noindent We employ the sigmoid function
\begin{equation}
	g : \mathbb R \to [0, 1] \ , \qquad g(x) = \frac{1}{1+e^{-x}}
\end{equation}
as activation function.
Hence, data propagate through the network as  
\begin{equation}
	a_{d}^{(\ell+1)} = g\left(\sum_{s=0}^{n_{\ell}} \Theta_{s, d}^{(\ell)} \cdot a_{s}^{(\ell)}\right), \quad 0 \leq d \leq n_{\ell+1}, \quad 0 \leq \ell < L \,
\end{equation}
being $\{a_{s}^{(0)},s=1,\ldots\}$  the input values. 

Specifically, at time $t$ our input values are the $(2\delta +1)A$ values obtained by restricting the matrix $\RSSI$ to the $\delta$ columns before and after $t$ (analogous notation to Section \ref{ssec:SW}), i.e. the column block $\RSSI_{\cdot,t-\delta\cdots t+\delta}$. The network output are $R$ numbers in $[0,1]$ that we interpret -- after $L^1$ normalization -- as the instantaneous probability of being in the $r$-th room.

We train the network parameters via gradient descent such that the network output fits hand-annotated data. 
In particular, we consider a dataset of $5427$ manually labelled input samples, $80\%$ of which are used to effectively set the weights, while the remaining  $20\%$ is employed for testing (i.e. checking for generalization and absence of overfitting).

\subsubsection{Estimating a trajectory via neural networks}
Applying the neural network to the $\RSSI$ matrix yields a $R \times T'$ ($T' = T-2\delta$) matrix $\mathcal P$ whose columns contain the probability of finding the considered visitor in room $r$ at time $t$.

Almost always, the network selects a room with an overall majority (probability $> 0.5$, see Figure~\ref{F:TrElaboration}\textbf e). However, it can 
happen -- particularly during a room transition -- that no class reaches such a majority. We apply therefore an \textit{ad hoc} adjacency filter:
probabilities values smaller than a fixed threshold, $\chi = 0.15$, are removed from the candidates; then, unfeasible transitions are either penalised or removed (e.g.\ transitions that imply wall crossings or that cover more than two rooms).
After data re-normalization, in the unlikely event that no room is selected with an overall majority, we assume the visitor in the closest room or in the same room.

\subsubsection{Comparing with the sliding window approach}
Looking at Figure \ref{F:TrElaboration}\textbf{d} and Figure \ref{F:TrElaboration}\textbf{f} we observe two interesting features: the sliding window approach often prevents spikes from arising (see minutes $\sim 35, 53$), yielding cleaner trajectories that makes it easier to enumerate room transitions. 
On the other hand, it tends to smooth trajectories excessively, ignoring fast transitions (see minute $\sim 52$).

The latter phenomenon is verified if we consider the bin-by-bin accuracy, i.e.\ the ratio between the correct predictions and the total number of samples analyzed.
The accuracy achieved over the test set by the neural network is in fact $0.858$ compared to $0.734$ obtained by the sliding windows approach.
Both of them however overcome results obtained via the argmax approach, which has an accuracy of $0.547$.

\subsection{Handling not-detected beacons}\label{sec:notdetected}
Due to (small) areas uncovered by antennas or because of random signal losses, it may happen that a beacon remains not detected. This is also what (correctly) happens before and after a visit or when the visitor leaves the museum during the visit, for example to reach the toilet.

Before performing statistical analyses, we amend for  not-detected statuses whenever this can be done unambiguously.
Although we notice that such a process might require mainly museum-specific solutions, we report two corrections which we deem of general interest.
\begin{enumerate}
	\item
	If the ``blind period'' is less than 3 minutes (10 minutes for the Pinacoteque), and the visitor is detected in the same room before and after the blind period, the visitor is associated with that room for the whole period.
	\item 
	If the blind period is less than 30 seconds, and the visitor is detected in two different rooms before and after the blind period, the visitor is supposed to be in one between the two rooms (at random).
\end{enumerate}

Whenever the overall not-detected status exceeds 25 minutes, we remove such a trajectory from the dataset.
Performing such pre-processing on the measurements collected during our field campaign, we obtain a dataset of $N=848$ trajectories, which will be the object of the analysis in the next sections.

\section{Trajectory analysis and clustering} \label{sec:trajectoryanalysisandclustering}

In this section we analyze the two datasets (including 848 trajectories) reconstructed via the two methods described before.

\subsection{Basic statistics}
We consider three basic illustrative statistics - that we also employ in Section \ref{sec:model} to calibrate our digital twin:
\begin{description}
	\item[Time of Permanence]
	    we denote by ToP$(v,r)$ the total time spent by visitor $v\in\{1, \ldots, N\}$ in room $r\in\{1, \ldots, R\}$ during their visit. 
	\item[Returning visitors]
    	we denote by RET$(v,r)$ the number of times visitor $v$ stopped by room $r$.
	\item[People per Room]
    	we denote by PpR$(r,t)$ the number of visitors in room $r\in\{1, \ldots, R\}$ during the time bin $t\in\{1, \ldots, T\}$. 
\end{description}
Note that the first two indicators are Lagrangian, while the third is Eulerian.

To perform such statistics we employ the trajectories as reconstructed by the neural network (Section \ref{MLapproach}) since its accuracy is higher and spurious spikes do not affect the analysis.

\subsubsection{Size of the dataset}
Before using the datasets and the indicators introduced above, we need to check if the number of considered trajectories is enough to get useful information. To this end, we have compared ToP and PpR extracted from the whole dataset (848 trajectories) with ToP and PpR extracted from several random subsets of the datasets with variable size. We observed that the difference between partial and complete datasets stabilizes starting from 300 sampled trajectories, meaning that we would not have obtained significantly different results if we had tracked more than 300 trajectories.

\subsubsection{Time of Permanence} \label{sec:top}

\begin{figure}[t]
	\centering
	\includegraphics[width=0.9\linewidth]{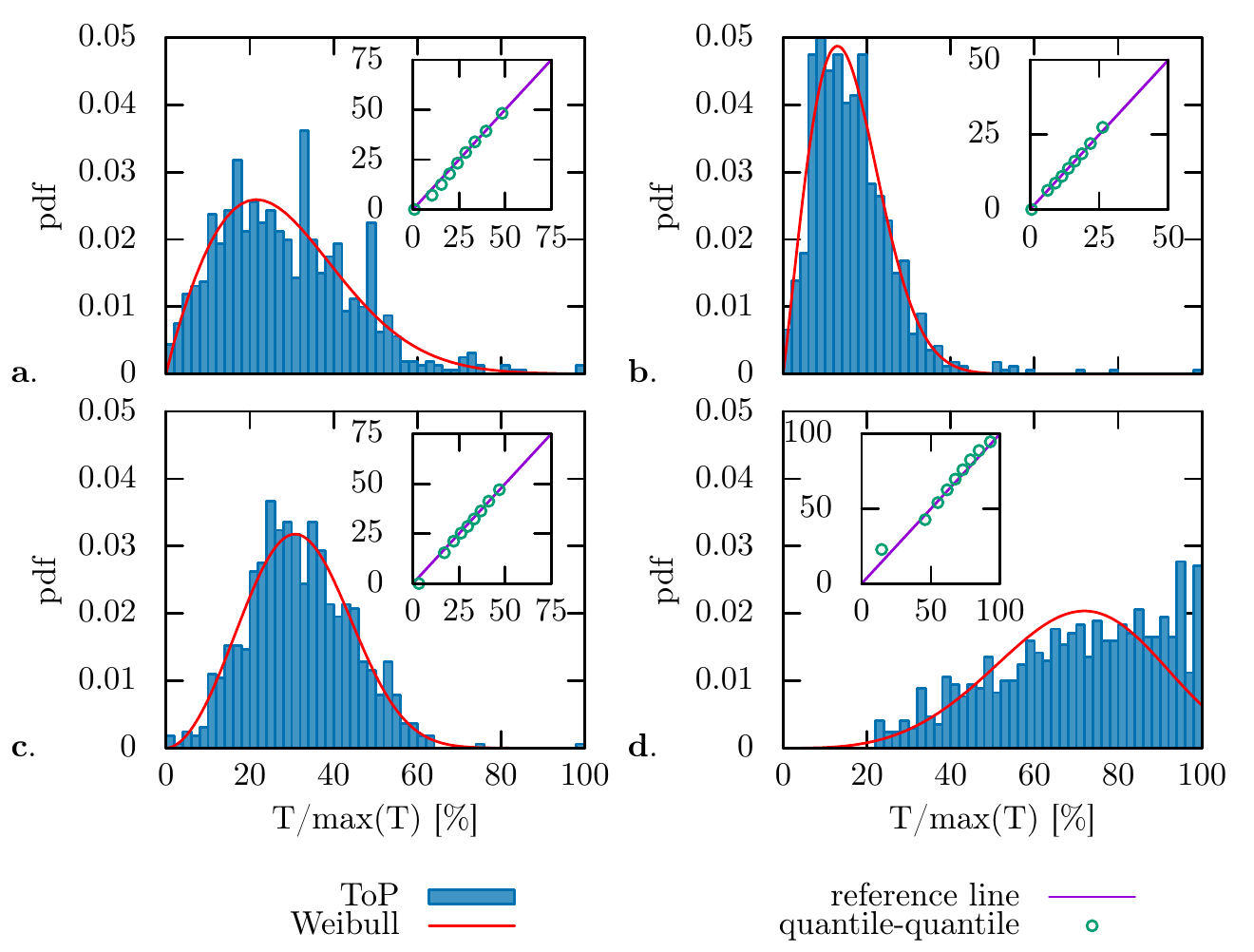}
	\caption{
		Four ToP distributions and their Weibull fit.
		\textbf{a.} \emph{Satiro su delfino} ($r_7$, $k=1.8$, $\lambda=17$).
		\textbf{b.} \emph{Apollo e Dafne} ($r_3$, $k=2$, $\lambda=36$). 
		\textbf{c.} Pinacoteque ($r_9$, $k=2.8$, $\lambda=221$).
		\textbf{d.} Whole museum ($k=4.1$, $\lambda=572$). 
		In the last case, the Weibull distribution does not fit correctly due to the forced exit after 2h. This problem will be solved later in Section \ref{sec:optim_no_bell}, by censoring the last 5 minutes of the visit.
		Inset: related Quantile-Quantile plots that depict the Real vs.\ Weibull quantile relation.
	}
	\label{F:wei}
\end{figure}

For each room $r$, the distribution of $\{$ToP$(v,r)\}_{v\in\{1,\ldots,N\}}$ is well fit by a Weibull distribution with $r$-dependent parameters $\lambda$ and $k$. 
The Weibull distribution performed best among the tested ones according to the Akaike Information Criterion. In Figure \ref{F:wei}, we report ToP empirical distributions and their Weibull fit for selected individual rooms as well as for the whole museum.

The Weibull distribution is related to the ``time-to-failure'' of a system, which, in our context, is to be interpreted as the ``time-to-exit'' a room (more precisely as the ``time-to-exit-and-do-not-return'', since we consider the ToP as the total time spent in a room).
The parameter $\lambda$ (characteristic time of visit) gives information about the room holding power, while parameter $k$ (Weibull slope) characterizes the decision to leave the room.
In particular, for all rooms of Galleria Borghese $k>1$ holds, meaning that the exit (failure) rate increases with time ($k=1$ indicates that the exit rate is constant over time while $k<1$ indicates a decreasing-in-time exit rate). 
In practice, we deem that visitors find all the rooms worthy of attention, and it  happens rarely that visitors leave a room immediately. 

In Section \ref{sec:model}, the survival and the hazard function associated to the Weibull distributions will be used as building blocks of the proposed museum digital twin.

\subsubsection{Returning visitors}
Galleria Borghese has a circular structure with no fixed or suggested path for visitors (cf.\ Section \ref{sec:casestudy}). 
Besides, the density of artworks is so remarkably high that visitors easily miss a fraction of the pieces during the first passage of a room.
Therefore, we investigate the number of times a person visits the same room, on average. 
In this analysis, we neglect quick returns (less than a minute of permanence). 

We observe that each guest visits a room, on  average, 1.3 times (1.5 times, for Room 8, \emph{Caravaggio}), while entrance rooms have 2.7 passages. 
On the other hand, 25\% of the visitors skips at least one room (especially room 7, \emph{Satiro su delfino}). The time of permanence during the first passage by a room is generally the longest, in comparison to the next ones (this, however, does not hold for entry rooms). 
The time of first return  (time interval between the moment a visitor leaves a room and the moment they return) appears consistent throughout the museum rooms and is between 25 and 30 minutes.

Finally, we highlight that the occurrence of fast returns (less than 5 minutes), which in our case are about 10\% of all returns, could indicate that visitors frequently get lost or change the direction of visit (clockwise vs.\ counterclockwise, cf.\  museum map in Figure \ref{fig:GBmap}).

\subsubsection{People per room}
\label{sec:ppr}
The number of people per room, PpR, is probably the most relevant indicator as well as that of largest interest for museum curators, as it connects with safety (hyper-congestion), comfort, and attractiveness for the audience (under-used rooms could indicate scarce interest).
We calculate the PpR$(r,t)$ by counting the number of visitors of each turn who are in room $r$ in time bin $t$. 
To amend for the fact that we gave beacons to a sample of visitors, and only to one member of each social group, we consistently replicate each trajectory $q$ times, where the integer $q$ is uniformly distributed between 1 and 6. 
We compare the PpR time series of one room and of the whole museum with our simulations in Section \ref{sec:simulation-results}.

\subsection{Measuring the distance among trajectories: a Wasserstein-inspired metric}
Defining a suitable metric in the space of trajectories is an essential step to quantify how `close' (or `similar') are distinct paths followed by visitors.
We define a new, \emph{ad hoc}, metric inspired by the Wasserstein distance, which is usually employed to quantify the distance between two abstract measures or two density functions. Following the same ideas, we quantify how much it costs to transform one visitor trajectory, time bin-by-time bin, into another, until they are identical.

First of all, we model the map of the museum as a graph with $R$ nodes. Edges between pairs of nodes represent viable connections between rooms. 
We also consider the possibility that the museum is organized in different \emph{wings}: wings are independent areas, which are so far from each other that it is natural to assume that visitors rarely visit the same wing twice (it is usually the case when a museum has multiple floors, comes in different buildings, or has multiple thematic areas). 
For technical reason, we always assume that there exists a wing called `Out'. Visitors in this wing are waiting to enter or have already left the museum. Thus, we represent Galleria Borghese in three wings: main floor, Pinacoteque, and Out.

Second, we introduce a distance function \emph{on the graph}, i.e.\ between room-nodes.
We define the distance $\mathcal D(r^1,r^2)$ between room $r^1$ and room $r^2$ ($r^1,r^2 \in \{1, \ldots, R\}$) by
\begin{equation}\label{eq:defD}
 \mathcal D(r^1,r^2):=
 \left\{
 \begin{array}{ll}
     0, & r^1=r^2, \\
     -\frac\alpha 2+\alpha\mathcal T_r(r^1,r^2)+\beta \mathcal T_w(r^1,r^2), & r^1\neq r^2,
 \end{array}
 \right. 
\end{equation}
where $\mathcal T_r(r^1,r^2)$ is the minimum number of room transitions (i.e.\ graph edges to hop through) necessary to traverse the graph from room $r^1$ to room $r^2$, whereas $\mathcal T_w(r^1,r^2)$ is the number of wing transitions, and $\alpha, \ \beta >0$ are two parameters. 
The term $-\frac\alpha 2$ is motivated by the need of decreasing the weight of short transitions (each room transition counts $\alpha$ but the first one which, instead, counts $\frac\alpha 2$), which often happen as many visitors stand still at the interface/door between two rooms, without actually moving in either direction.
Note that the distance $\mathcal D$ is not necessarily commutative. This can happen e.g.\ if a museum comes with some one-way room transitions.

Third, all trajectories are extended in order to have the same number of time bins.
We achieve this by fixing a maximal theoretical duration of the visit (2h in our case) and then exploiting the fictitious wing Out, where people are placed before and after the actual visit.
In this case, we have performed the analysis using the sliding window approach (Section \ref{ssec:SW}). This choice was driven by the need of lessening the amount of spikes, see, e.g., Figures \ref{F:TrElaboration}\textbf{d} and \ref{F:TrElaboration}\textbf{f} at minute $\sim 35$ where spikes may arise between rooms located on different floors, affecting the reliability of the measure.

\begin{figure}[t]
	\begin{center}
            \includegraphics[width=0.60\linewidth]{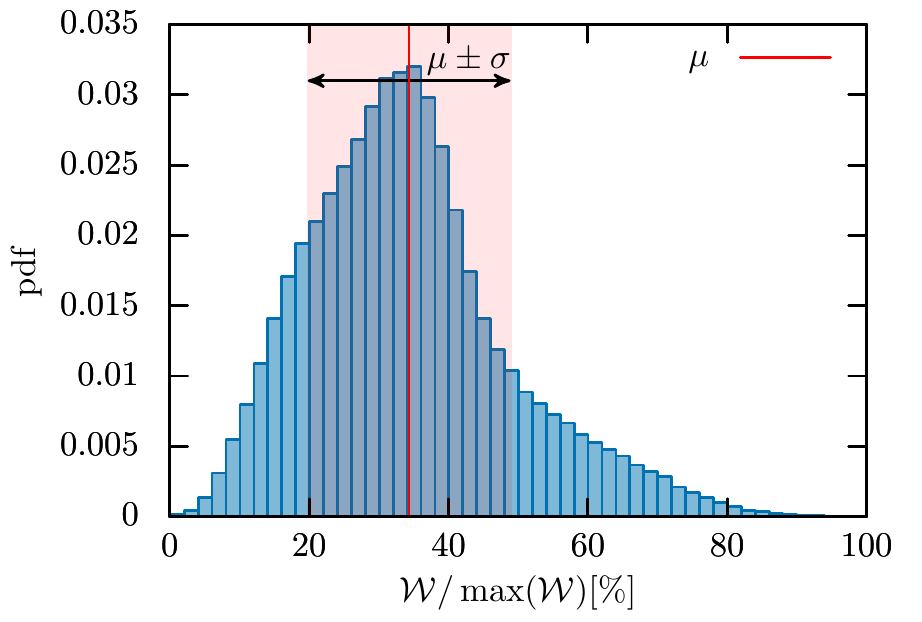}
		\caption{
			Distribution of mutual distances between trajectories. The $x$-axis is normalized w.r.t.\ the longest measured distance.
			The mean pairwise distance $\mu$ is reported in red while the shaded area denotes the range $\mu \pm \sigma$ ($\sigma$ being the standard deviation of the distribution).
		}
		\label{F:likelihood}
	\end{center}
\end{figure}

We have now all the ingredients to define the distance between two trajectories $\textsc{t}^1,\textsc{t}^2\in\{1,\ldots,R\}^T$
\begin{equation}\label{eq:defW}
    \mathcal W(\textsc{t}^1,\textsc{t}^2):=\sum_{t=1}^T \mathcal D(\textsc{t}^1_t,\textsc{t}^2_t),
\end{equation}
which represents the sum, time bin-by-time bin, of the distances between the rooms (according to \eqref{eq:defD}) occupied by the two visitors at each given instant. 
As an example, in Figure \ref{F:likelihood} we report the distribution of the pairwise distances between all trajectories, computed using metric \eqref{eq:defW}.

\begin{figure}[t]
	\begin{center}
		\includegraphics[width=0.9\linewidth]{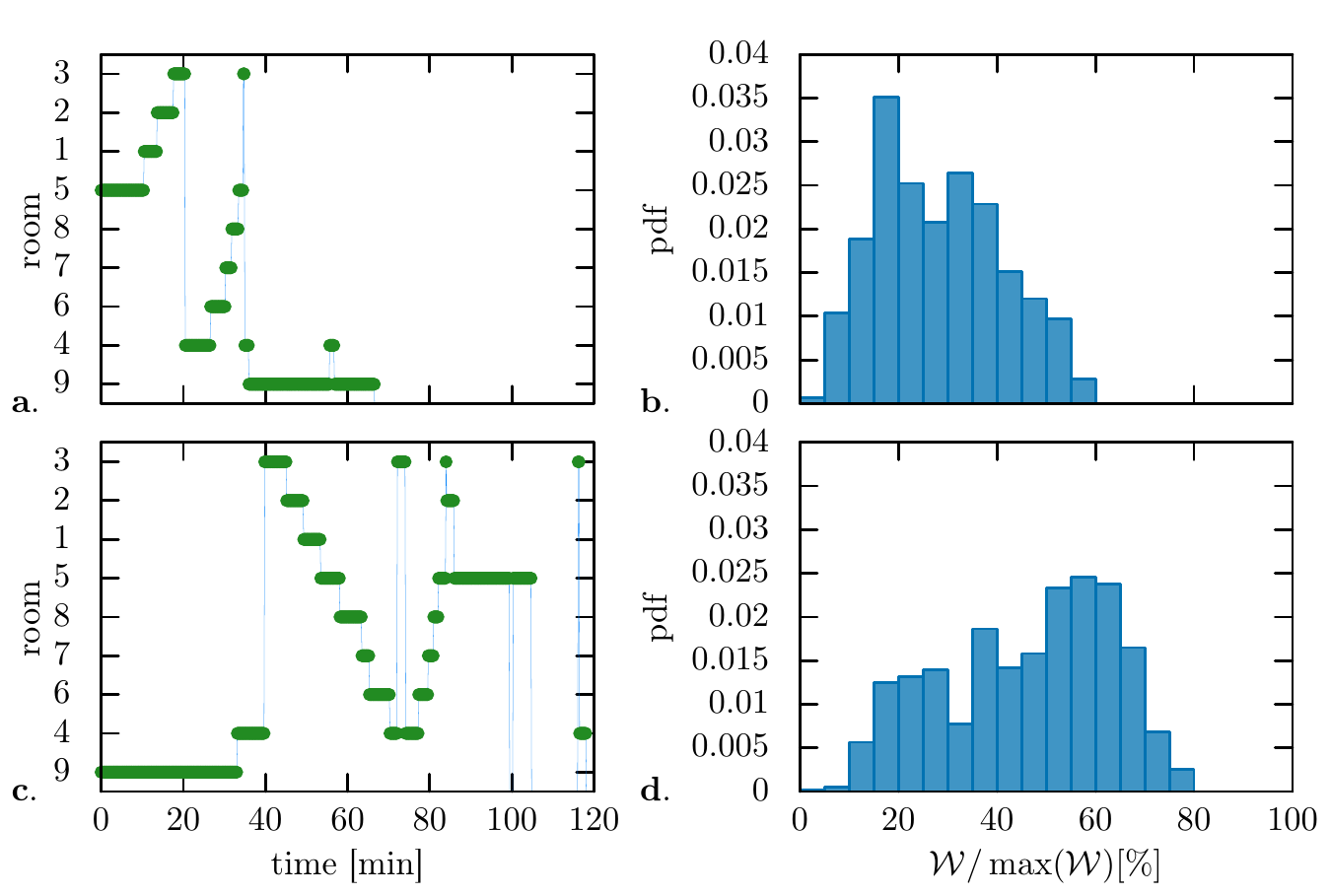}
		\caption{
			\textbf{a.} Most common trajectory and, \textbf{b.}, distribution of the distances between such trajectory and all the others. The visitor performs a circular visit following the room numbering in the main floor, then they reach the Pinacoteque upstairs.
			\textbf{c.}\ \& \textbf{d.} Analogous plot for the least common trajectory in our dataset. The visitor enters the museum via the Pinacoteque, then they visit the main floor twice, once clockwise and once counterclockwise.
			$x$-axis in \textbf{b.}\ \& \textbf{d.}\ is normalized w.r.t.\ the longest measured distance among all the trajectories.
		}
		\label{F:most-least-common}
	\end{center}
\end{figure}

Figure \ref{F:most-least-common} shows instead the most and least common trajectory. The higher the number of trajectories `close' to a given one, the more common the trajectory is. Hence, we report the single trajectory having the highest number of other trajectories within distance $\mu -\sigma$ (Figure \ref{F:most-least-common}\textbf{ab}, cf.\ definition of $\mu$ and $\sigma$ in Figure \ref{F:likelihood}) and  the single trajectory having the least number of other trajectories within distance $\mu + \sigma$ (Figure \ref{F:most-least-common}\textbf{cd}).

\begin{figure}[t]
	\begin{center}
		\includegraphics[width=0.9\linewidth]{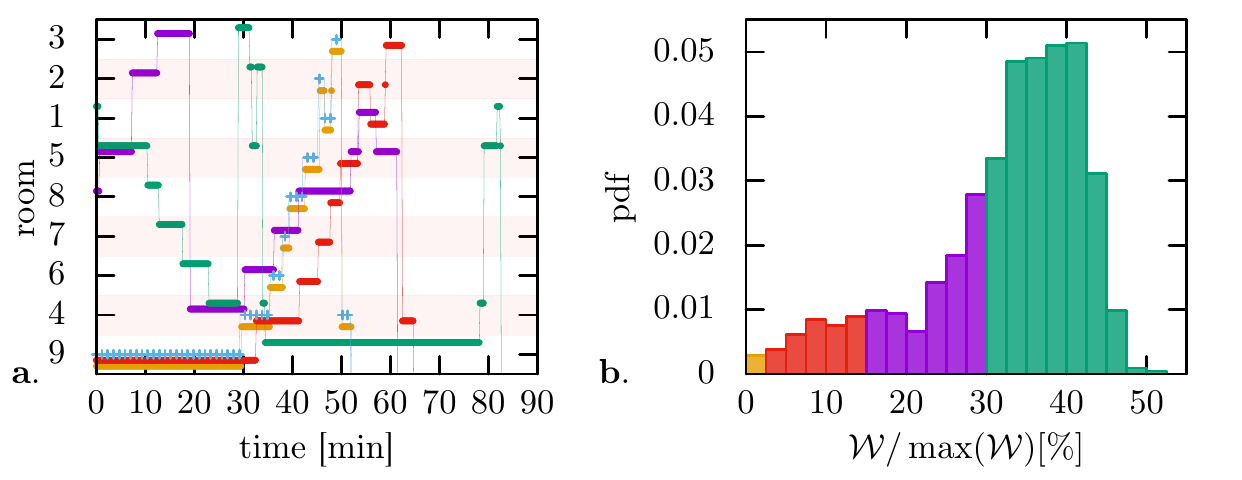}
		\caption{
			\textbf{a.} Sample of measured trajectories. 
			\textbf{b.} Distribution of the distances between the trajectory marked by ``blue plus signs'' ({\color{Cerulean}$+$}) in \textbf{a.} and all the others.
			Distances are reported in percentage w.r.t.\ the longest distance measured.
			Trajectories closer than $0.025\%$ (yellow bin) likely belong to the same group of visitors;
			Trajectories closer than $0.15\%$ (red bins) are slightly time shifted; 
			In trajectories closer than $0.30\%$ (purple bins) relations are still identifiable;
			Trajectories farther away than $0.30\%$ (green ones) are completely unrelated. 
			Trajectories in \textbf{a.} are random sampled from corresponding color percentile sets in \textbf{b.}
		}
		\label{fig:group-visitors_0}
	\end{center}
\end{figure}

Finally, we mention the capability of finding automatically members of social groups (it could happen that elements of the same social group went to the ticket office separately, thus were assigned more than one beacon). Indeed, two or more trajectories very close to one another likely belong to visitors in company. Figure \ref{fig:group-visitors_0} reports a sample of trajectories at different distance from a given reference trajectory.

\subsection{Clustering algorithms}\label{sec:clustering-algorithms}
As we recalled in Section \ref{sec:refs}, clustering  algorithms can be used for inferring, from the  whole trajectories data set, the typical paths or, equivalently, the typical individual behaviors inside the museum. 

Here we employ algorithms which do not require to define \emph{a priori} the number $k$ of clusters, nor to assign predefined reference trajectories around which clusters are agglomerated (as typically happens with, e.g., $k$-means approaches). 
Moreover, we do not use the typical taxonomy (ant, butterfly, fish, grasshopper, cf.\ Section \ref{sec:refs}) to guide the clustering, aiming at other, possibly hybrid, behaviors.
To this end, we employ an \emph{agglomerative hierarchical clustering} (AHC) approach (see, e.g., \cite{duda2012pattern}). 
These techniques consider a bottom-up cluster tree (dendrogram), that, step by step, gathers trajectories according to their mutual likelihood.
In the beginning, each trajectory is considered to be the only element of a distinct cluster. Then, at each iteration, the two closest clusters get merged into a single cluster.
The process is deterministic, unless we have two couples of clusters at exactly the same distance, and it always ends with one single cluster after $N-1$ steps, given $N$ the number of initial trajectories.
Note that cutting the dendrogram at the $\ell$-th layer from the tree leaves provides exactly $N-\ell$ clusters.
Finding an adequate cutting layer is an issue which we discuss in the following.

To measure the distance between two clusters, we leverage on \eqref{eq:defW}. 
We consider, in particular, three common methods:
\begin{description}
	\item[C-LINK]
		in Complete Linkage, the distance between two clusters $\mathcal C^1$ and $\mathcal C^2$ is the maximum amongst the distances between all the trajectories within the two clusters:
		\begin{equation}
			\mathcal W(\mathcal C^1, \mathcal C^2) = \max\{\mathcal W(\textsc{t}^1, \textsc{t}^2) : \textsc{t}^1 \in \mathcal C^1, \textsc{t}^2 \in \mathcal C^2\}.
		\end{equation}
	\item[S-LINK]
		in Single Linkage, the distance between two clusters $\mathcal C^1$ and $\mathcal C^2$ is the minimum amongst the distances between all the trajectories of the two clusters:
		\begin{equation}
			\mathcal W(\mathcal C^1, \mathcal C^2) = \min\{\mathcal W(\textsc{t}^1, \textsc{t}^2) : \textsc{t}^1 \in \mathcal C^1, \textsc{t}^2 \in \mathcal C^2\}.
		\end{equation}
	\item[UPGMC]
		in Unweighted Pair Group Method with Mean Centroid, each cluster $\mathcal C$ is identified by a representative trajectory $\bar{\textsc{t}}_{\mathcal C}$, and the distance between two clusters is evaluated as the distance between representative trajectories:
		\begin{equation}
			\mathcal W(\mathcal C^1, \mathcal C^2) = 
			\mathcal W(\bar{\textsc{t}}_{\mathcal C^1},\bar{\textsc{t}}_{\mathcal C^2}).
		\end{equation}
\end{description}

Determining a representative trajectory $\bar{\textsc{t}}_{\mathcal C}$ in a trajectory set $\mathcal C$ is useful in general, and mandatory to employ UPGMC.
To do so, we compute a mode among all the trajectories: for each time bin $t$, our representative trajectory reports the most visited room among the elements of $\mathcal C$:
\begin{equation}\label{E:centroid}
	(\bar{\textsc{t}}_{\mathcal C})_t :=
	\arg\max\limits_{r \in \{1,\ldots, R\}}\left\{
		\sum_{\textsc t \in \mathcal C} \one_{\textsc t_t = r}
	\right\}\ ,
	\qquad 0 < t \leq T\ .
\end{equation}
Note that the centroids found with a specific cut may also be employed to clusterize a different set of trajectories.
This also means that, if new trajectories are gathered, the same centroids may be used in order to get a clustering. 
This may reveal that habits have been broken or new paths have been discovered.

\medskip

\emph{Cutting the dendrogram.} 
In order to find the right cutting threshold for the dendrogram, we consider the number of the $p$-\emph{significant} clusters, i.e.\ the clusters with more than $p$ elements, while traversing the tree from the leaves to the root.
Having a high variation in the number of significant clusters in the proximity of the root often implies that clusters are unstable, i.e.\ they merge randomly in the process, preventing valuable interpretations. Having instead a very small number of significant clusters, say one or two, often means that each cluster contains very nonhomogeneous elements, thus resulting practically useless for categorization. 

\begin{figure}[t]
	\centering
	\includegraphics[width=0.97\linewidth]{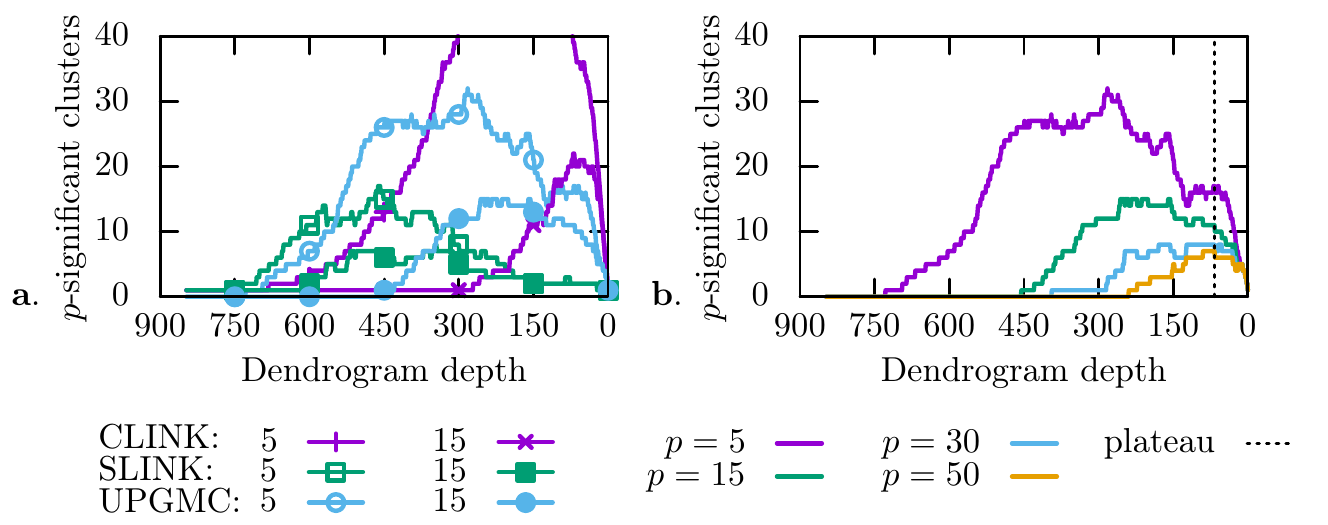}
	\caption{
		\textbf{a.} Number of 5- (filled markers) and 15- (empty markers) significant clusters as a function of the dendrogram depth for C-LINK, S-LINK and UPGMC methods.
		\textbf{b.} Number of $p = 5, 15, 30, 50$ significant clusters obtained via UPGMC method.
		The dendrogram is cut in correspondence to the plateau at depth $67$.}
	\label{F:dendrogram-density}
\end{figure}

Figure \ref{F:dendrogram-density}\textbf a reports the number of 5- and 15-significant clusters as a function of the dendrogram depth, for the three methods described before.
C-LINK yields many small unstable clusters joining together, with no meaningful interpretation, towards the end of the process. 
S-LINK offers, on the other hand, a poor set of typical clusters to which all the trajectories converge quickly throughout the clustering process.
Conversely, UPGMC leads to a relatively small amount of consistent stable clusters.

In particular, the UPGMC dendrogram shows a plateau around layer $\bar\ell\sim 67$, for many values of $p$, see Figure \ref{F:dendrogram-density}\textbf b. We adopt such a cutting layer since it ensures the maximum amount of highly significant clusters ($p = 30, 50$ have the last absolute maximum there) without trading-off too much information in smaller clusters.

\subsection{Clustering results} \label{sec:clustering-results}
We consider here the representative trajectories of each cluster obtained after a dendrogram cut at layer $\bar\ell$.
Although none of the representative trajectories strictly coincides with any among the trajectories observed, they all appear real (i.e.\ conform with a potential visit). This emphasizes that clusters indeed aggregate similar trajectories.

\begin{figure}[t]
	\begin{center}
		\includegraphics[width=0.95\linewidth]{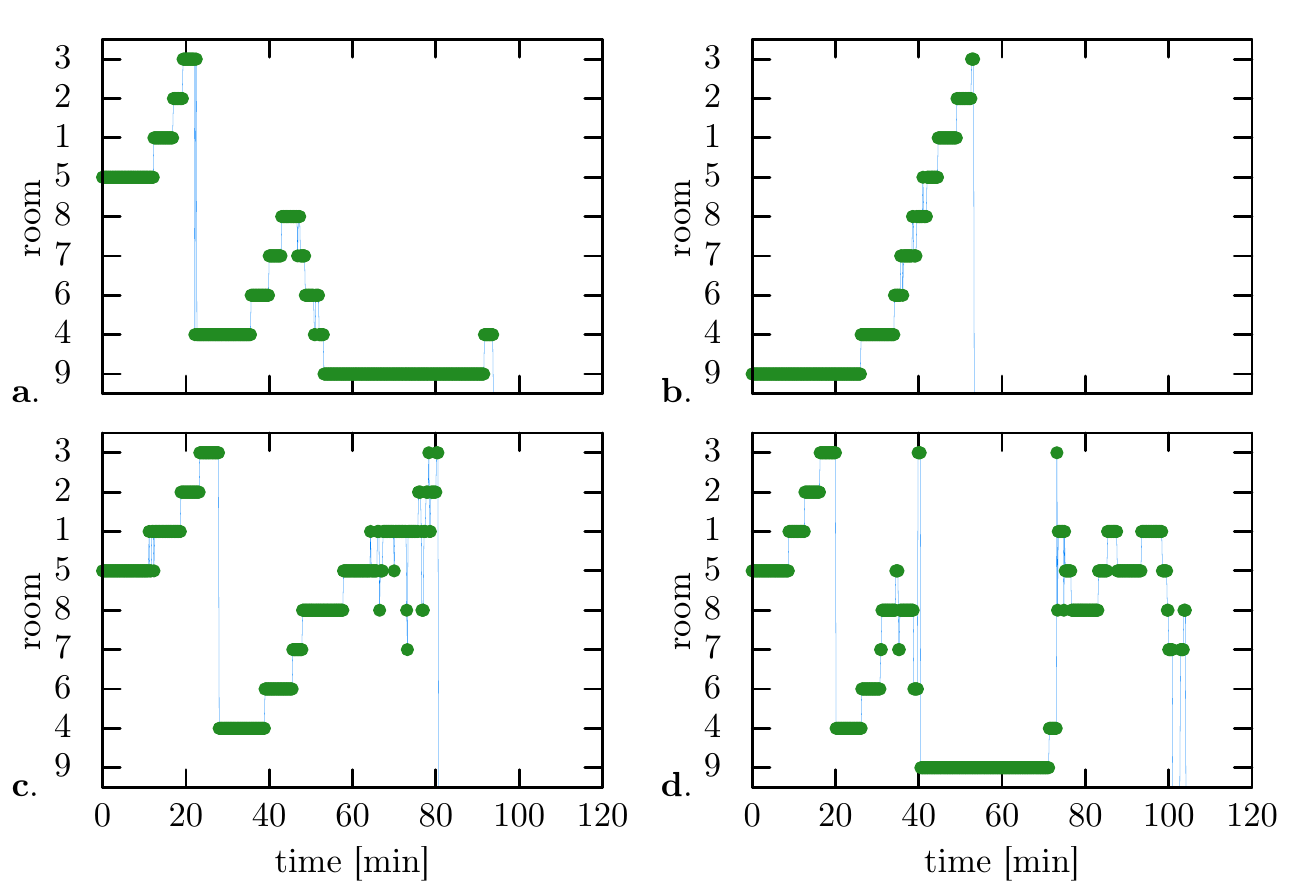}
	\end{center}
	\caption{
		Four representative trajectories (centroids) of clusters joining respectively \textbf{a.} 16\%, \textbf{b.} 9\%, \textbf{c.} 4\%, \textbf{d.} 1\% of the trajectory data set.
		Representative trajectories may show spikes (see, e.g., \textbf c., $\sim 75$ minute). According to \eqref{E:centroid}, this phenomenon arises whenever rooms have approximately the same number of visitors within the same interval of time.
		}
	\label{fig:clusters}
\end{figure}

Figure \ref{fig:clusters} shows four representative trajectories related to four clusters of different size.
The two most common patterns are related to visitors who follow the natural numbering of the rooms, starting or ending the visit in the Pinacoteque, which is visited once.
This identifies the most typical visit pattern for the curators. 
Nevertheless, clustering investigation brings to light other, less expected, patterns: the one which does not include the visit at the Pinacoteque (possibly visitors who did not find the staircase) and patterns where the Pinacoteque comes amidst the visit. 
Note that both patterns have been observed by the museum managers and are discouraged. 

\emph{Filtering by clustering.} Clustering can be also used to detect unfeasible/unreal trajectories coming from system malfunctioning, since those trajectories tend to gather in a single cluster. 
This powerful feature helps to design filters to clean up the data during the preprocessing phase.

\begin{figure}[t]
	\begin{center}
		\includegraphics[width=0.95\linewidth]{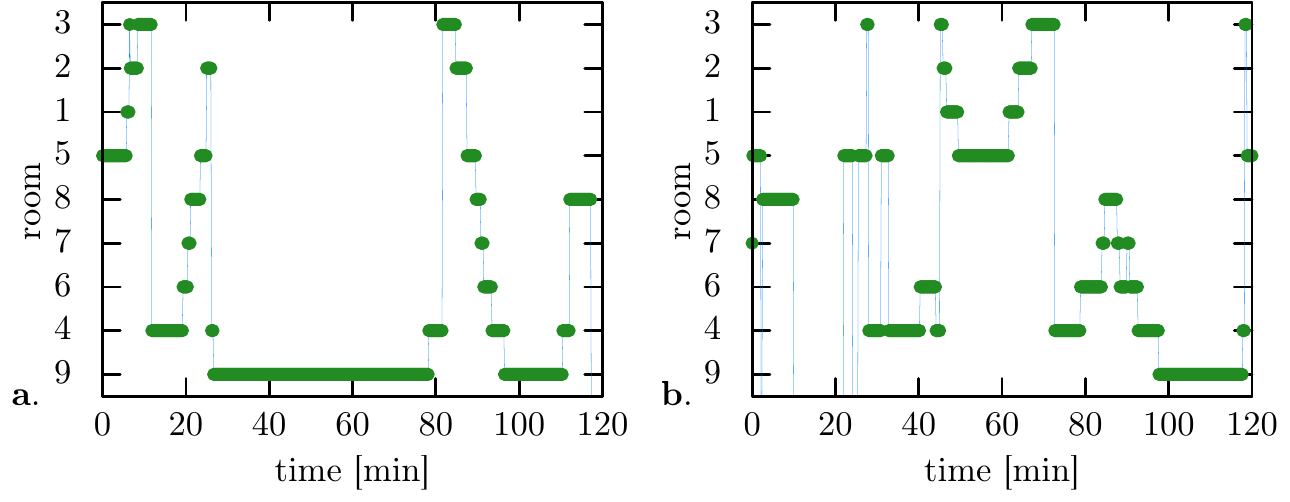}
	\end{center}
	\caption{
		Two anomalies detected.
		\textbf{a.} A rare pattern where the Pinacoteque and main floor are both visited twice. 
		\textbf{b.} A strange pattern with many changes of direction (clockwise/counterclockwise).
	}
	\label{F:anomalies}
\end{figure}

\emph{Anomaly detection.} Trajectories which remain isolated  in the last layers (close to the root of the tree) are, by definition, far from all the other centroids and therefore very atypical.
We claim that these trajectories are \emph{anomalies} detected during the process. 
If they do not come from system malfunctioning, they belong to people who behave abnormally or suspiciously and deem additional checks.
Figure \ref{F:anomalies} shows some of the anomalies detected in our study.

\section{Model and calibration}\label{sec:model}
In this section, we develop a digital twin of the museum, i.e.\ an algorithm which is capable of generating new trajectories, (statistically) indistinguishable from measured ones.

In order to represent the complex visitor behaviour, we employ a stochastic approach based on Markov Chains (MC). We design our simulator to generate visiting paths with relevant observable features such as guests skipping one or more rooms and/or returning multiple times to the same room.

\medskip

The model is based on two important assumptions:
\begin{description}
\item[Visitors are independent from each other]
the decision to leave or remain in a room does not depend on the number of people in that room.
This assumption is certainly reasonable up to mild congestion levels. On the other hand, hyper-congestion has surely an impact on visitors choices, however, our current data collection seems still insufficient to quantify such a challenging aspect.
We suspect that congestion can either increase or decrease the ToP, depending on the perceived importance and fame of the room content.
\item[Social groups behave as one individual]
social groups visit the museum remaining together, i.e.\ following the same trajectory and thus spending the same time in each room. 
This assumption, which is an important limitation, is consistent with the fact that beacons were given almost always to a single member of each social group. 
Therefore, we are not capable of disentangling interactions and differences within social groups.
\end{description}

In a standard MC, the transition probability from a state (room) to the next depends only on the current state. However, in our context, it is an intuitive idea that the visitors choices depend, in some way, on the rooms that they have previously visited. Furthermore, since we are assuming that there is no predefined visit path, a standard MC creates a \emph{bounce phenomenon} among rooms, (i.e.\ $1 \to 2 \to 3 \to 2 \to 1 \to 2$), while the majority of paths are more regular (e.g.\ $1 \to 2 \to 3 \to 4$ or $4 \to 3 \to 2 \to 1$).

In order to simulate the complex visitor behavior, we introduce a \emph{memory} in the Markov Chain, to represent  the visitors knowledge of the visited rooms. Moreover,  we reasonably assume that visitors also remember the time spent in each room.
We use a nonhomogeneous transition matrix, which is time-dependent through a weight function $S$. Henceforth, we refer to this model as Time Varying Markov Model.

\subsection{Time Varying Markov Model (TVMM)} \label{sec:TVMM}
\begin{figure}[t]
\centering
\scalebox{0.7}{

\definecolor{white}{RGB}{255,255,255}
\definecolor{ccccccc}{RGB}{204,204,204}
\definecolor{cffcc00}{RGB}{255,204,0}
\definecolor{c99cc00}{RGB}{153,204,0}

\begin{tikzpicture}[draw=black,fill=cffcc00]

	
	\path[fill=ccccccc, draw] (-4.,2.0) rectangle ++(3.0,3.0);
	\path[fill=ccccccc, draw] (-4.,5.0) rectangle ++(1.25,0.5) node[pos=.5] {Out};
	\path[fill=ccccccc, draw] (0.0,0.0) rectangle ++(7.0,7.0);
	\path[fill=ccccccc, draw] (0.0,7.0) rectangle ++(1.5,0.5) node[pos=.5] {Wing 1};
	\path[fill=ccccccc, draw] (8.0,1.0) rectangle ++(5.0,5.0);
	\path[fill=ccccccc, draw] (8.0,6.0) rectangle ++(1.5,0.5) node[pos=.5] {Wing 2};
	
	
	\path[fill, draw] (5.5,0.5) rectangle ++(1.0,1.0) node[pos=.5] {$r_1$};
	\path[fill, draw] (5.5,3.0) rectangle ++(1.0,1.0) node[pos=.5] {$r_2$};
	\path[fill, draw] (5.5,5.5) rectangle ++(1.0,1.0) node[pos=.5] {$r_3$};
	\path[fill, draw] (3.0,4.5) rectangle ++(1.0,2.0) node[pos=.5] {$r_4$};
	\path[fill, draw] (3.0,0.5) rectangle ++(1.0,2.0) node[pos=.5] {$r_5$};
	\path[fill, draw] (0.5,5.5) rectangle ++(1.0,1.0) node[pos=.5] {$r_6$};
	\path[fill, draw] (0.5,3.0) rectangle ++(1.0,1.0) node[pos=.5] {$r_7$};
	\path[fill, draw] (0.5,0.5) rectangle ++(1.0,1.0) node[pos=.5] {$r_8$};
	\path[fill, draw] (8.5,1.5) rectangle ++(4.0,4.0) node[pos=.5] {$r_9$};
	\path[fill=c99cc00, draw] (-3.5,2.5) rectangle ++(2.0,2.0) node[pos=.5] {$r_{\mbox{\footnotesize out}}$};
	
	
	\draw[->, line width=3.05pt] (5.75,1.50) -- (5.75,3.00) node [pos=.5, above, rotate=90] {0.61};
	\draw[->, line width=1.95pt] (5.50,0.75) -- (4.00,0.75) node [pos=.5, below] {0.39};
	
	\draw[->, line width=0.85pt] (6.25,3.00) -- (6.25,1.50) node [pos=.5, above, rotate=-90] {0.17};
	\draw[->, line width=3.80pt] (5.75,4.00) -- (5.75,5.50) node [pos=.5, above, rotate=90] {0.76};
	\draw[->, line width=0.40pt] (5.50,3.75) -- (4.00,5.25) node [pos=.5, below, rotate=-45] {0.08};
	
	\draw[->, line width=1.00pt] (6.25,5.50) -- (6.25,4.00) node [pos=.5, above, rotate=-90] {0.20};
	\draw[->, line width=4.00pt] (5.50,5.75) -- (4.00,5.75) node [pos=.5, below] {0.80};
	
	\draw[->, line width=1.05pt] (4.00,6.25) -- (5.50,6.25) node [pos=.5, above] {0.21};
	\draw[->, line width=1.80pt] (3.00,5.75) -- (1.50,5.75) node [pos=.5, below] {0.36};
	\draw[->, line width=2.15pt] (3.75,6.50) -- (4.50,7.75) -- (9.5,7.75) node [pos=.5, below] {0.43} -- (10.0,6.00);
	\draw[->, line width=2.5pt, dashed] (3.25,6.50) -- (2.50,8.00) -- (-1.00,8.00) node [pos=.5, above] {$h(t)$} -- (-2.25,5.00);
	
	\draw[->, line width=2.80pt] (4.00,1.25) -- (5.50,1.25) node [pos=.5, above] {0.56};
	\draw[->, line width=0.85pt] (4.00,1.50) -- (5.50,3.25) node [pos=.5, above, rotate=45] {0.17};
	\draw[->, line width=1.35pt] (3.00,0.75) -- (1.50,0.75) node [pos=.5, below] {0.27};
	\draw[->, line width=2.50pt, dashed] (3.50,0.50) -- (3.50,-0.5) -- (-2.5,-0.5) node [pos=.5, below] {$h(t)$} -- (-2.5,2.00);
	
	\draw[->, line width=0.80pt] (1.50,6.25) -- (3.00,6.25) node [pos=.5, above] {0.16};
	\draw[->, line width=4.20pt] (1.25,5.50) -- (1.25,4.00) node [pos=.5, above, rotate=-90] {0.84};
	
	\draw[->, line width=0.45pt] (0.75,4.00) -- (0.75,5.50) node [pos=.5, above, rotate=90] {0.09};
	\draw[->, line width=4.55pt] (1.25,3.00) -- (1.25,1.50) node [pos=.5, above, rotate=-90] {0.91};
	
	\draw[->, line width=4.45pt] (1.50,1.25) -- (3.00,1.25) node [pos=.5, above] {0.89};
	\draw[->, line width=0.55pt] (0.75,1.50) -- (0.75,3.00) node [pos=.5, above, rotate=90] {0.11};
	
	\draw[->, line width=5.00pt] (10.25,6.0) -- (9.75,8.50) -- (4.25,8.50) node [pos=.5, above] {1.00} -- (3.50,6.50);
	\draw[->, line width=2.50pt, dashed] (10.5,6.00) -- (10.5,9.25) -- (-2.5,9.25) node [pos=.5, above] {$h(t)$} -- (-2.5,5.00);

\end{tikzpicture}}
\caption{
	Transition probabilities between rooms in Galleria Borghese (the probability to remain in the same room is not included).
	We can see that the counterclockwise path is preferred and fast transitions from rooms 5, 2 to rooms 2, 4, respectively, exist.
	The probability of leaving the museum is sampled as a hazard function $h(t)$.
}
\label{F:TM}
\end{figure}
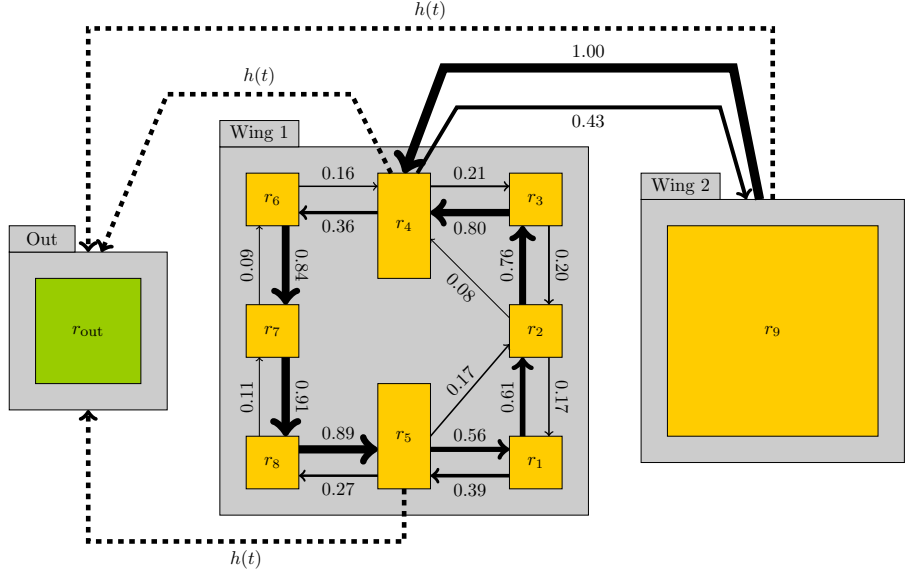

Since the museum comes with $R$ rooms, we consider a $R\times R$ transition matrix $\mathcal K$. Following the frequentist definition of probability, $\mathcal K_{r^1,r^2}$ is computed by first counting, from all the measured trajectories, the number of transitions from room $r^1$ to room $r^2$, where $r^1=r^2$ holds if the visitor remains in the same room.
\begin{equation}\label{eq:counter}
	\mathcal K_{r^1,r^2}=\sum_{n=1}^{N} k_{n}(r^1,r^2),
\end{equation}
where $k_n(r^1,r^2)$ denotes the number of $r^1\rightarrow r^2$ transitions along the $n$-th trajectory of the data set. 
The sum over columns of $\mathcal K_{r^1,r^2}$ represents the total time, in time bin, spent by all tracked visitors in room $r^2$.
If we normalize $\mathcal K$ by row, so that $\sum_{r^2} p_{r^1,r^2}=1$, we obtain a transition matrix $\mathcal M$ where the new element $p_{r^1,r^2}$ represents the probability to move from room $r^1$ to room $r^2$, see Figure \ref{F:TM}.

In order to avoid the room bouncing phenomenon, we make the transition matrix $\mathcal M$ time-dependent. More precisely, we consider the matrix
\begin{equation}\label{eq:transition-matrix-intro}
	\tilde{\mathcal M}_{r^1,r^2}(t)=\mathcal M_{r^1,r^2} \ S_{r^2}(t),\qquad r^1,r^2 \in \{1,\ldots,R\},
\end{equation}
where $S_r(t)$ is the survival function associated to ToP$(\cdot,r)$ via its Weibull fit parameters $(\lambda_r,k_r$),
\begin{equation}
S_r(t) =  e^{-(t/\lambda_r)^{k_r}}.
\end{equation}
In other words, $S_r(t)$ quantifies the probability that a guest visits room $r$ for a time interval longer than $t$.
$S_r(t)$ is a decreasing function such that $S_r(0)=1$ and $S_r(t_{max})=0$, where $t_{max}$ is the largest measured ToP$(\cdot,r)$.

At each time step of the simulation, the function $S_r(t)$ must be updated on the basis of the time spent in each room, and the transition matrix $\tilde{\mathcal M}$ has to be normalized by rows in order to have a correct definition of the transition probability.

In the following, we detail the exceptions to the transition dynamics in \eqref{eq:transition-matrix-intro} to cope with the access and exit conditions.

\begin{description}
\item[Beginning of a visit] 
in Galleria Borghese visitors enter all together at the beginning of the visit turn. However, due to some delay (ticket control, late arrival, queue at wardrobe), the entrance process is completed in about 20 minutes. We simulate these dynamics extracting the delay at random from the set of measured delay. In addition, we use another probability distribution function to assign the entrance room (we recall that Galleria Borghese has three entrances: \emph{Ratto di Proserpina} (room 4), \emph{Portico} (room 5), and Pinacoteque (room 9)).
\item[Conclusion of a visit] 
the wing Out is conceptually different from the other wings. Therefore, we manage the exit time in a distinct manner. The exit is not controlled by $\tilde{\mathcal M}$, instead it is managed via the hazard function $h$ of the Weibull distribution with parameters $(\lambda_{*},k_{*})$ associated to the total time of visit, see Figure \ref{F:wei}\textbf d. 
More precisely, at every time bin $t$ of the simulation, the exit probability is given by
\begin{equation}\label{eq:hazardexit}
	P(r_{t+1}=\text{Out}\mid r_{t}=r_{exit})=h(t;k_{*},\lambda_{*}),
\end{equation}
where $r_{exit}$ is an exit room and
\begin{equation}
	h(t;k_{*},\lambda_{*})=\frac{k_{*}}{\lambda_*^{k_{*}}}t^{k_{*}-1}.
\end{equation}
\end{description}

\subsection{Simulation results} \label{sec:simulation-results}
Before presenting the results of our model, we explain how we compare simulated trajectories with the measurements in our data set. We perform such comparison to quantify the accuracy of the simulation.
The observables of interest are the following:
\begin{description}
	\item [ToP]
	we evaluate the mean and the coefficient of variation of the ToP distributions, for both real and simulated visits. 
	\item [PpR] 
	we consider 100 statistically independent simulated turns, each including a total of 400 generated trajectories (similarly to Section \ref{sec:ppr}, simulated visits are replicated $q$ times, where $q$ is a uniform integer random variable between $1$ and $6$, to mimic social groups). We compute the PpR at each time bin as an ensemble average across such 100 realizations.	
	\item [Clusters] 
	we use the same clustering technique presented in Section \ref{sec:clustering-algorithms} to aggregate simulated trajectories.
	This analysis aims to check if the most numerous cluster is sufficiently close to the measurements; this guarantees that the simulator creates a sufficient amount of plausible trajectories.
\end{description}

\begin{figure}[t]
	\begin{center}
		\includegraphics[width=0.9\linewidth]{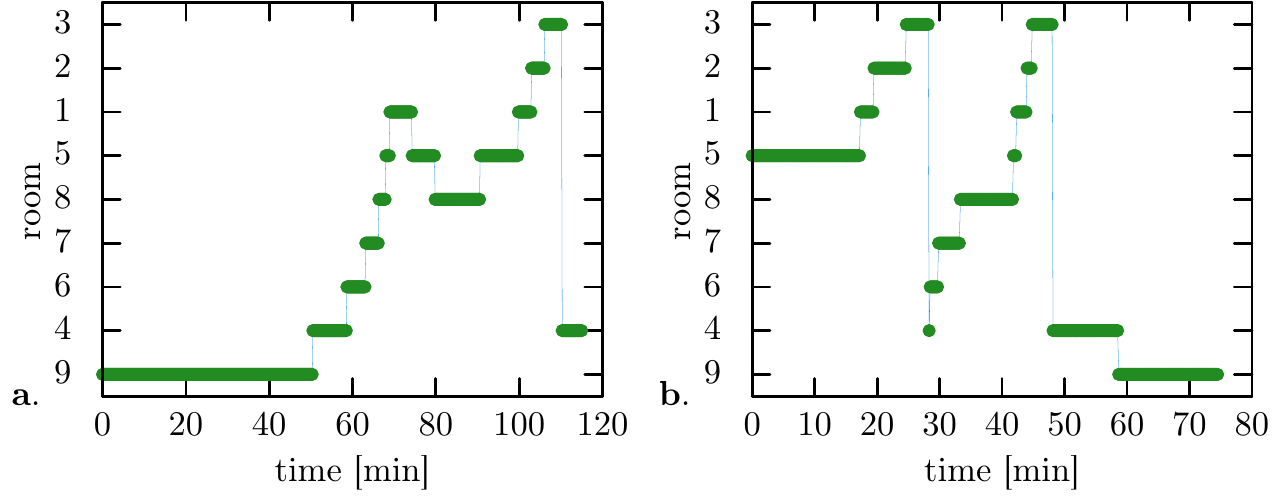}
		\caption{Two simulated trajectories: \textbf{a.} A long trajectory which begins from the Pinacoteque (room 9) and then moves to the main floor according the room enumeration. 
		\textbf{b.} A short trajectory that begins from room 5, traverses the main floor according to the room enumeration, and finally reaches the Pinacoteque. 
		}
		\label{fig:simulatedtrajectories}
	\end{center}
\end{figure}

Figure \ref{fig:simulatedtrajectories} shows two simulated trajectories, which indeed share typical features with measurements: the Pinacoteque is visited once and the visit path follows the natural numbering of rooms. At times, people come back to rooms already visited, as in real life. 

Table \ref{T:delta} compares the real and simulated ToP distributions, by considering the relative differences in ToP averages ($\mu$) and variation coefficient ($\mbox{VC}=\sigma/\mu$, $\sigma$ being the standard deviation of the ToP distribution), respectively 
\begin{equation}
\delta \mu = \frac{\mu_{sim}}{\mu_{real}} - 1 \qquad \delta \mbox{VC} = \frac{\mbox{VC}_{sim}}{\mbox{VC}_{real}} - 1.
\end{equation}
The mean values of the distributions are well approximated, despite the simulations tend to slightly overestimate the ToP in the main floor and to underestimate it in the Pinacoteque. The $\delta$VC indicator highlights instead some differences between model and data: real visitors are more unpredictable than simulated ones, which yields negative $\delta$VC values. On the contrary, the dynamics in the Pinacoteque appears predictable and even more consistent than in simulations. This  most likely relates to the fact that the Pinacoteque is the area with the weakest antenna coverage: amending not-detected data diminishes the variance of the measured ToP distribution. 

\begin{table}[t]
	\begin{center}
	\begin{tabular}{c c c c c c c c c c c}
	    \hline
        {\small Room } &   {\small 1} &  {\small 2} & {\small 3} & {\small 4} & {\small 5} & {\small 6} & {\small 7} & {\small 8} & {\small 9} & {\small Museum} \\ 
        \hline\hline
        {\small $\delta \mu$ } & $11\%$ & $10\%$ & $2\%$ & $8\%$ & $12\%$ & $-3\%$ &	$-2\%$ & $7\%$ & $-8\%$ & $-3\%$ \\
        {\small $\delta$VC } & $-28\%$ & $-10\%$ & $-15\%$ & $-15\%$ & $-20\%$ & $13\%$ & $12\%$ & $3\%$ & $31\%$ & $-11\%$  \\
        \hline
    \end{tabular}
    \caption{
        Relative error between mean and coefficient of variation of ToP distribution evaluated for real trajectories and simulated ones.
        $\delta x= (x_{\mbox{\footnotesize sim}}/x_{\mbox{\footnotesize real}}-1)$, where $x$ is either $\mu$ or $VC$.
    }
    \label{T:delta}
	\end{center}
\end{table}

\begin{figure}[t]
	\begin{center}
		\includegraphics[width=0.9\linewidth]{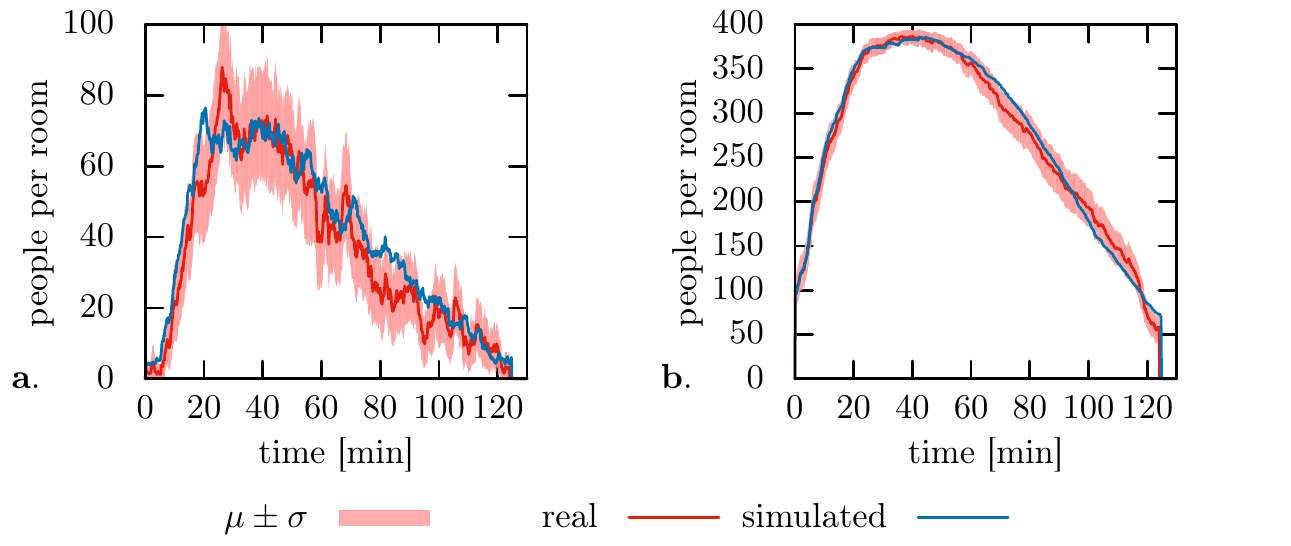}
		\caption{
		Comparison of the average PpR of real visits (red line) and the ensemble-average PpR of simulated visits (blue line) in \textbf{a.} \emph{Ratto di Proserpina} and in \textbf{b.} the whole museum.
		The shaded area corresponds to the interval $[\mu-\sigma,\mu+\sigma]$. We note that the blue line is almost entirely contained in the shaded area, as expected. 
		}
		\label{fig:comparePpR}
	\end{center}
\end{figure}

In Figure \ref{fig:comparePpR}, we compare measurements and simulations considering the PpR as a function of time. Simulations are reported in terms of ensemble statistics among 100 realizations, in particular we consider ensemble PpR average and ensemble PpR standard deviation. 

\begin{figure}[t]
	\begin{center}
		 \includegraphics[width=0.45\linewidth]{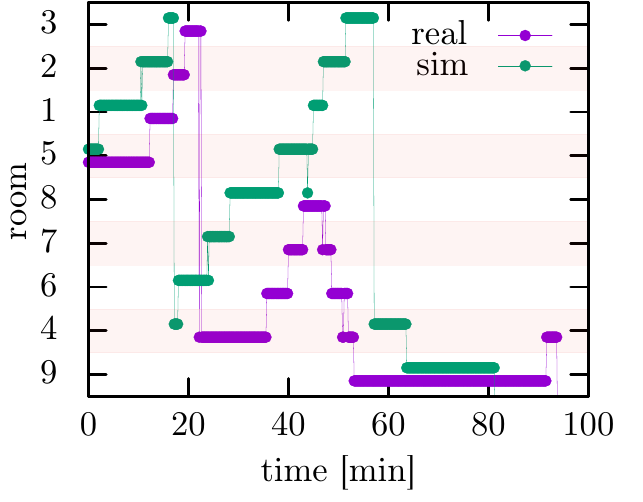}
		\caption{
		The two most numerous clusters obtained gathering real and simulated trajectories.
		The real case joins $16\%$ of real trajectories, whereas the simulated one $18\%$. 
		We note that they share a number features, e.g., the ToP in each room, the total time of visit, the entry room (\emph{Portico}, room 5), and the final room (Pinacotque, room 9). The main difference is the behavior after completing the visit on the main floor. Real visitors come back counterclockwise, while simulated visitors keep walking clockwise. 
		This could be explained by the fact that many visitors ask for information in room 5 and are sent backwards to the staircases. The model does not include the interactions with the museum staff, hence cannot catch this feature.
        }
		\label{fig:compareclusters}
	\end{center}
\end{figure}

In Figure \ref{fig:compareclusters} we finally report the representative trajectories of the two most numerous clusters obtained by gathering real and simulated trajectories. 
\medskip

\section{Museum control and optimization} \label{sec:optimization}
We are now ready to employ the digital twin introduced in the previous section as a tool to improve the museum experience. More precisely, we simulate different scenarios and observe visitors behavior in virtual environments, aiming at supporting curators decisions. 
In this regard, it is useful to remark that changing the ticketing strategy or the duties of security staff can require weeks of training in real life.

For our case study, we identify the following control variables and objectives. \\
\emph{Control variables:}
\begin{description}
    \item[C1] considering that Galleria Borghese has three entrances (\emph{Ratto di Proserpina}, \emph{Portico}, Pinacoteque), museum managers can assign a certain percentage of visitors to each entrance (currently they are 15\%, 60\%, and 25\% respectively). 
    Operationally, such control can be implemented by introducing a tag (e.g.\ name or color) in the ticket which specifies the entrance.
    \item[C2] the scheduled entry times in the museum can be tuned (currently visitors enter at 09:00, 11:00, 13:00, 15:00, and 17:00, after the museum empties).
    \item[C3] the number of visitors allowed in each turn can be modified (currently it is 360 reserved in advance plus 30 last-minute).
    \item[C4] the fixed duration of a visit turn can be either (C4a) modified or (C4b) totally removed (currently it is set to 2h slots).
\end{description}
\emph{Objectives:}
\begin{description}
    \item[O1] keeping the PpR below a certain room-dependent threshold. Historically, our study began precisely to control the number of visitors in the Pinacoteque, which has a very low admittance limit for safety reasons.
    \item[O2] keeping the PpR, in any room at any time, approximately constant. This would reduce strong variations of relative humidity which can damage the artworks \cite[Chapter 2]{camuffo1998book}.
    \item[O3] decreasing the queue at the entrance. 
    \item[O4] increasing the number of visitors per day.
\end{description}

Note that O1, considering the emergency situation caused by the COVID-19 virus pandemic, can be employed to respect the imposed social distances legislation.

Among the many possibilities, we focused on two improvements: C1 aiming at O1, and C2, C3 \& C4b aiming at O1 \& O2.

\subsection{Entrance strategy optimization}
Keeping the existing conditions regarding the number of visitors and the time horizon, we explore the effects of a different visitor partition among the three entrances (C1). 
We aim at a PpR as low as possible in all rooms (O1), especially in the Pinacoteque, which is the room with the most stringent safety constraints.

We fix an \emph{overcrowding threshold} for each room, representing a PpR limit the curators do not want to exceed.
Then, we pursue a brute force attack to the optimization problem, trying all the possible triplets $(E_1, E_2, E_3)\in [0,100]^3$, $\sum_{e=1}^3 E_e = 100$, which indicate the percentage of visitors starting the visit from each entrance $e=1$ (\emph{Ratto di Proserpina}), $e=2$ (\emph{Portico}), and $e=3$ (Pinacoteque). 

\begin{figure}[t]
	\centering
	\includegraphics[width=0.75\linewidth]{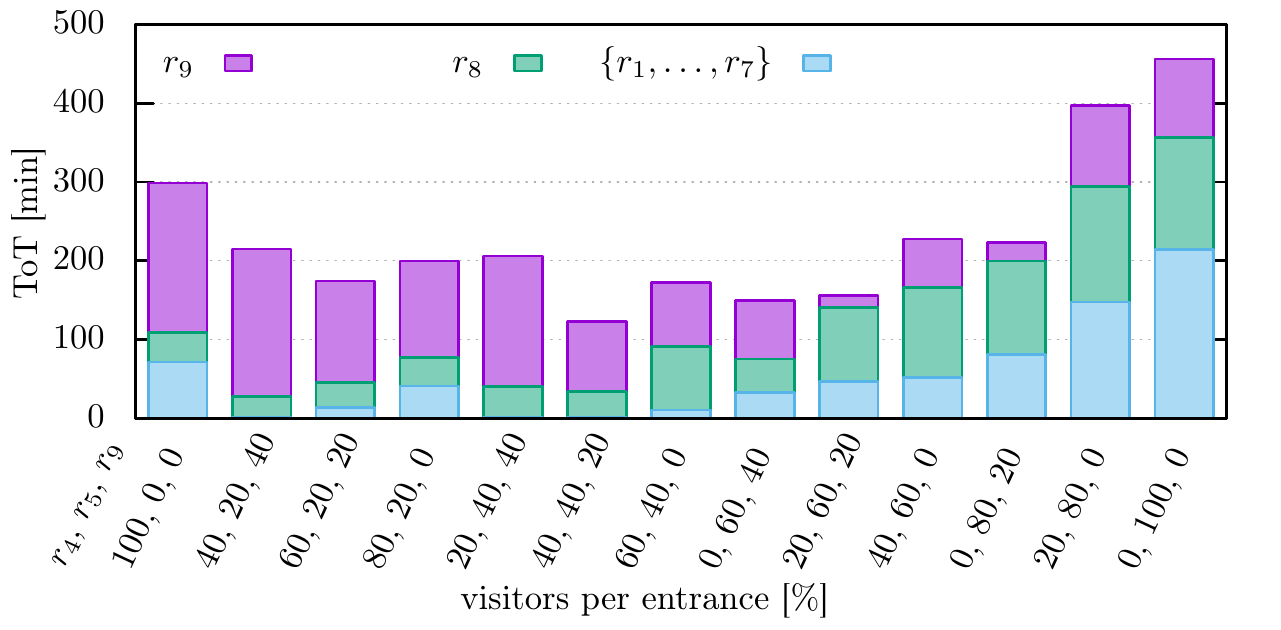}
	\caption{
	    Total time duration in which the overcrowding threshold is exceeded (ToT) in room 9 (Pinacoteque), in room 8 (\emph{Caravaggio}) and in all other rooms (sum of each ToT is considered), for 13  triplets $(E_1, E_2, E_3)$.
	    We observe that the overall ToT exceeds at least 120 min over a day of visit regardless of the entrance system.
	}
\label{fig:entranceoptimization}
\end{figure}

Figure \ref{fig:entranceoptimization} shows the results of the optimization process evaluating the \emph{total time the PpR exceeds the overcrowding threshold} (ToT), for room 9 (Pinacoteque), room 8 (\emph{Caravaggio}), and for all the remaining rooms.
The best triplet for the Pinacoteque is $(20,60,20)$, while the best triplet for \emph{Caravaggio} is $(40,20,40)$: in fact, these configurations minimize the ToT in those rooms, respectively.
More in general, it is easy to see that optimal choices for one room do not necessarily mean optimality for others. 
The solution currently employed by the museum, which is $(15,60,25)$, is almost optimal to reduce overcrowding in Pinacoteque and in the whole museum in general, but it sacrifices the pleasantness of the visit in some rooms of the main floor.

\subsection{Removing the finite time horizon of the visits}\label{sec:optim_no_bell}
The full elimination of the current finite time horizon allowed for the visits is a challenging improvement for the museum experience.
The idea is to keep the reservation mandatory, with entry interval fixed every 30, 60, or 120 minutes (C2), but, unlike the current setting, \emph{remove the requirement to leave after 2h} (C4b).
The immediate advantage is that the museum staff does not have to empty the museum at the end of the visit turn, thus saving about 5-7 minutes during which the museum remains completely empty (O2).
Moreover, this would also be a great advantage for the (few) visitors who want to stay for a very long time inside the museum.

Unfortunately, as it happens for every mathematical model, simulation results are reliable only in the conditions in which the simulator was developed and calibrated.
In our measurements, less than 1/4 of visitors are still inside the museum when the time limit is reached (and are forced to exit);  for these a (negative) influence of the time limit certainly occurs.
Nevertheless, such influence possibly exists also for the other 3/4, that might have scheduled their visit according to the existing time constraints.

\begin{figure}[t]
	\centering
   \includegraphics[width=0.9\linewidth]{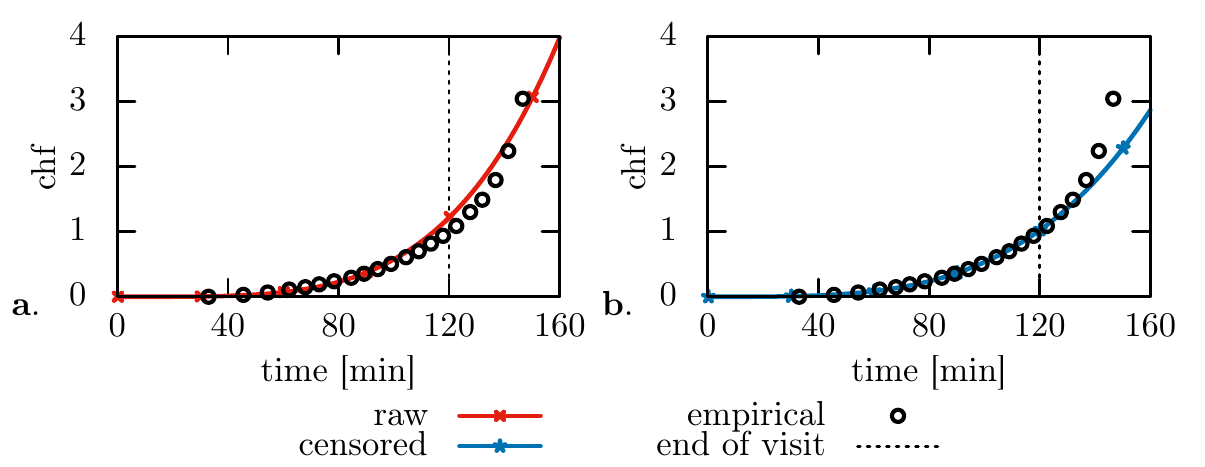}
	\caption{
	Cumulative hazard function associated to the Weibull distribution of the whole museum.
	Empirical values are calculated with the Kaplan--Meier method.
	\textbf{a.} Without censoring (cf.\ Figure \ref{F:wei}\textbf{d.}) and
	\textbf{b.} after censoring the last 5 minutes of visit (new parameters are $k_{*}=3.5$ and $\lambda_{*}=596$). 
	This method allows us to get a better fit of the real distribution between 0 and 2h, i.e.\ the visit interval. The uncensored fit, instead, is negatively influenced by the forced exit. 
	}
	\label{F:weibull_censored}
\end{figure}

We attacked the problem by \emph{censoring} the Weibull distribution of the time of visit of the whole museum (cf.\ Section \ref{sec:top}). This statistical procedure allows us to deal with data set in which the event of interest is not observed during the study.
We obtain the new distribution as a maximum likelihood estimate censoring the last 5 minutes of visit, see Figure \ref{F:weibull_censored}.
We use the estimated parameters to modify the hazard function which controls the conclusion of the visit. 

\begin{figure}[t]
	\begin{center}
		\includegraphics[width=0.75\linewidth]{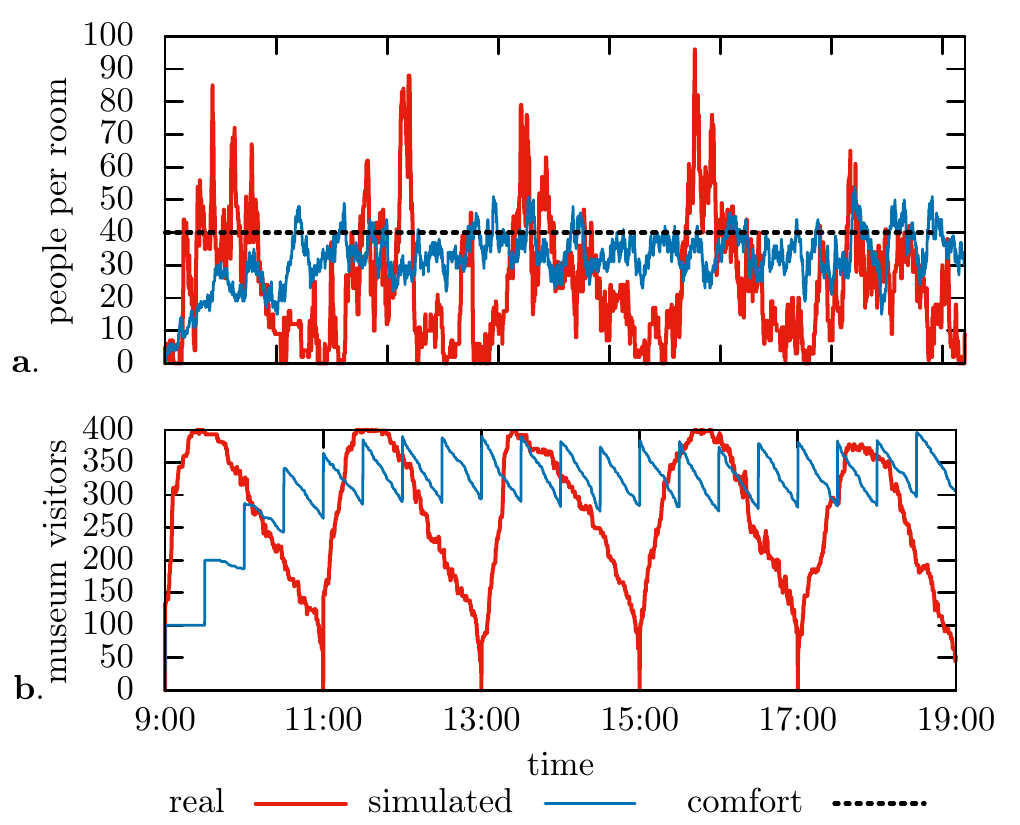}
		\caption{
		    PpR as a function of time in the current settings and considering the best entrance strategy. The comparison includes \textbf{a.} room 8 (\textit{Caravaggio}) and \textbf{b.} the whole museum.
        }
		\label{fig:removingbell}
	\end{center}
\end{figure}

We simulated an entire day i.e.\ 9 a.m. -- 7 p.m., corresponding to the total time span of the 5 visit turns currently implemented. 
This is necessary as after removing the time limit, visit turns overlap and the museum never empties. 
Figure \ref{fig:removingbell} shows the result of the optimization process. The best strategy is to let 100 visitors (C3) enter from the main floor (C1) every 30 minutes (C2). These choices eliminate completely the peaks in the PpR indicator (congestion moments, O1) and the PpR remains stable with small fluctuations during the whole visit day (O2). Having the system approximately at this thermodynamic-like equilibrium greatly facilitates the management since it allows to calculate -- using the measured transition matrix -- the average number of people in each room from the number of visitors allowed (i.e.\ sold tickets).

\section{Conclusions and future work} \label{sec:conclusions}
This study aimed at measuring, analyzing, modeling, and optimizing visitors behavior in museums and similar environments. 
The practical goal was to provide suggestions to museum curators for efficiently managing visitors flows. 

The implemented measurement system is sustainable for the museum, being economically viable and well accepted by visitors. 
A free application to be installed on the smartphone could serve as a beacon as well, provided visitors find it useful (as an audio-guide, for example). 
Employing Raspberry Pi's as fixed Bluetooth antennas appeared quite convenient and allowed the necessary development flexibility. 

A major issue surely comes from the noisiness of the Bluetooth signal, which must be overcome by suitable data post-processing. The sliding window approach has proven to be more effective in measuring room transitions, while the machine learning approach performed better at estimating the permanence time in the various rooms. 

From the trajectory analysis, we have identified some issues in the museum design and visit experience that can be considered by curators: for example, rooms of the same size have drastically different time of permanence, as it happens for \emph{Caravaggio} and \emph{Satiro su delfino}. 
This suggests that the museum can benefit from a rearrangement  of the artworks, although this is not always possible due to historical or architectural constraints. 
In addition, rooms like \emph{Paolina} have an uneven distribution of visitors, being congested in the first half of the visit turn and under-used in the second half.

The museum simulator allowed us to propose the implementation of a new ticketing and entrance system. 
The entry scheme identified is to let 100 people enter every 30 minutes from \emph{Portico} and \emph{Ratto di Proserpina} while eliminating the 2h time limit, thus reducing congestion and fluctuations of the number of people in each room. 

\medskip

In the next future, we plan to further improve the model presented here. In particular, we aim at including the internal dynamics of \emph{social groups} (families, friends, guided tours), and at considering the impact of congestion on individual behavior. This is to lift the current statistical independence of simulated trajectories, thus increasing the level of complexity.

The impact of visitors on the local microclimate is also an outstanding issue to which we aim. On the basis of the present work and \cite{desantoli2016}, one can achieve a coupled model for predicting temperature, humidity, and crowding, on which basis one can program intelligent air conditioning systems.

\section*{Acknowledgements}
We would like to thank Sara Suriano, Massimiliano Adamo, Federico Ricci Tersenghi, Elisabetta Giani, and all the staff of Galleria Borghese for all their time and support during this project.

\section*{Funding}
Results presented in this paper are achieved under the project \textit{Management of flow of visitors inside the Galleria Borghese in Rome}, supported by Ministry of Cultural Heritage and Activities and Tourism, Galleria Borghese, and Istituto per le Applicazioni del Calcolo of National Research Council of Italy.
Project's Principal Investigators are Marina Minozzi (Galleria Borghese) and Roberto Natalini (IAC-CNR). 

This work was also carried out within the research project ``SMARTOUR: Intelligent Platform for Tourism'' (No.\ SCN\_00166) funded by the Ministry of University and Research with the Regional Development Fund of European Union (PON Research and Competitiveness 2007-2013).

E. Cristiani also acknowledges the Italian Minister of Instruction, University and Research to support this research with funds coming from
the project entitled \textit{Innovative numerical methods for evolutionary partial differential equations and applications} (PRIN Project 2017, No.\ 2017KKJP4X).

A. Corbetta also acknowledges the support of the Talent Scheme (Veni) research programme, through project number 16771, which is financed by the Netherlands Organization for Scientific Research (NWO).

E. Cristiani is a member of the INdAM Research group GNCS.

\bibliographystyle{plain}
\bibliography{biblio_museums_official}
\end{document}